%% file: main.tex
\documentclass[10pt, aps, twocolumn, a4paper, prx, longbibliography, superscriptaddress]{revtex4-2}

\usepackage[utf8]{inputenc}
\usepackage[T1]{fontenc}
\usepackage{amsmath, amsthm, amssymb}
\usepackage{booktabs}
\usepackage{nicefrac}

\usepackage{graphicx}
\usepackage{physics}
\usepackage{enumerate}
\usepackage{array}

\usepackage[dvipsnames]{xcolor}
\definecolor{myblue}{RGB}{29, 113, 184}
\definecolor{myred}{RGB}{217, 61, 61}
\usepackage{hyperref}
\hypersetup{
	colorlinks=true,
	linkcolor=myred,
	filecolor=magenta,      
	urlcolor=myblue,citecolor=myblue}
\usepackage[capitalize]{cleveref}
\Crefname{appendix}{App.}{Apps.}
\usepackage{mathtools}

\usepackage[all]{hypcap} 

\newcommand{\br}[1]{\left( #1 \right)}
\newcommand{\bre}[1]{\left[ #1 \right]}
\newcommand{\brek}[1]{\left\{ #1 \right\}}
\DeclareMathOperator{\expecOp}{\mathbb{E}}
\newcommand{\Expec}[2]{\expecOp_{#1}\bre{#2}}
\DeclareMathOperator{\varOp}{\mathrm{Var}}
\newcommand{\Var}[2]{\varOp_{#1}\bre{#2}}

\newcommand{\integer}{\mathbb{Z}}

\newcommand{\reals}{\mathbb{R}}

\newcommand{\MyPaperTitle}{Synthesis of discrete-continuous quantum circuits with multimodal diffusion models}

\begin{document}

\title{\MyPaperTitle}
\author{Florian Fürrutter}
\affiliation{Department of Theoretical Physics, University of Innsbruck}
\author{Zohim Chandani}
\affiliation{Quantum Algorithm Engineering, NVIDIA Corporation}
\author{Ikko Hamamura}
\affiliation{Quantum Algorithm Engineering, NVIDIA Corporation}
\author{Hans J.~Briegel}
\affiliation{Department of Theoretical Physics, University of Innsbruck}
\author{Gorka Mu\~noz-Gil}
\email{gorka(dot)munoz-gil(at)uibk(dot)ac(dot)at}
\affiliation{Department of Theoretical Physics, University of Innsbruck}

\begin{abstract}
Efficiently compiling quantum operations remains a major bottleneck in scaling quantum computing. Today's state-of-the-art methods achieve low compilation error by combining search algorithms with gradient-based parameter optimization, but they incur long runtimes and require multiple calls to quantum hardware or expensive classical simulations, making their scaling prohibitive. Recently, machine-learning models have emerged as an alternative, though they are currently restricted to discrete gate sets. Here, we introduce a multimodal denoising diffusion model that simultaneously generates a circuit's structure and its continuous parameters for compiling a target unitary. It leverages two independent diffusion processes, one for discrete gate selection and one for parameter prediction. We benchmark the model over different experiments, analyzing the method's accuracy across varying qubit counts and circuit depths, showcasing the ability of the method to outperform existing approaches in gate counts and under noisy conditions. Additionally, we show that a simple post-optimization scheme allows us to significantly improve the generated ansätze. Finally, by exploiting its rapid circuit generation, we create large datasets of circuits for particular operations and use these to extract valuable heuristics that can help us discover new insights into quantum circuit synthesis. Code and models are publicly available at \href{https://github.com/FlorianFuerrutter/genQC}{https://github.com/FlorianFuerrutter/genQC}.
\end{abstract}

\maketitle 

\section{Introduction} \label{sec:introduction}
Synthesizing quantum circuits for given quantum operations is a highly non-trivial task, and stands as one of the key challenges in the pursuit of large-scale quantum computation. As quantum hardware continues to improve, with increasing qubit counts and lower error rates, we move closer to regimes where quantum advantage may become feasible. Indeed, quantum computational advantage has now been widely demonstrated, as for instance in Shor’s factoring algorithm~\cite{shor1994algorithms} and Grover’s search~\cite{grover1996fast}, or in applications such as optimization~\cite{munoz2025scaling} and machine learning~\cite{liu2021rigorous}. However, many of these algorithms rely on large, fault-tolerant quantum computers, which remain out of reach. Despite notable advances in hardware, we are still in the era of Noisy Intermediate-Scale Quantum (NISQ) devices~\cite{preskill2018quantum}, where limited qubit connectivity and different error sources hinder quantum computation.
Another practical challenge arises from the diversity of quantum computing paradigms, from gate-based quantum computers realized by photonic~\cite{madsen2022quantum}, superconducting~\cite{arute2019quantum}, neutral atoms~\cite{bluvstein2024logical} and trapped ions~\cite{monz2016realization} architectures to measurement-based quantum computation~\cite{briegel2009measurement}. While this variety is promising and enriches the field, it also introduces difficulties: each platform features different native gate sets and qubit connectivity constraints, meaning that the optimal compilation of a given operation can vary significantly from one architecture to another.

To address these challenges, various powerful compilation techniques have been developed, typically involving complex pipelines composed of multiple modules~\cite{smith2023LEAP, younis2021qfast, madden2022best,grimsley2019adaptive}. While they are able to output highly accurate circuits, these methods tend to be slow and rely on heuristics, search algorithms and gradient-based optimization, often resulting in runtimes that scale prohibitively with qubit count~\cite{davis2020towards, madden2022best}. Nonetheless, from a fundamental and theoretical perspective, improving circuit synthesis depends on a deeper understanding of quantum circuits and how different gate combinations give rise to different computations. Beyond a purely theoretical study, achieving this requires methods that can generate circuits not only accurately but also efficiently, enabling the creation of datasets from which, when combined with expert knowledge in quantum information and computation, better compilation strategies can be discovered.

In this work, we build on recent advances in machine learning-assisted quantum circuit synthesis to develop such a model.  
Specifically, we leverage denoising diffusion models (DMs) to generate quantum circuits consisting of both discrete and parameterized gates that compile a given unitary operation. Our main contributions are as follows:

\begin{enumerate}
    \item We present a method that synthesizes input unitaries into quantum circuits, predicting simultaneously the circuit's structure (i.e., gate types and distribution) and its continuous parameters. It is based on a multimodal diffusion model that couples two independent diffusion processes: one addresses the discrete task of selecting gate types, while the other proposes the corresponding continuous gate parameters. 

    \item We propose a strategy to pre-learn noise schedules for the discrete data part of the diffusion, ensuring proper mixing of discrete classes throughout the forward diffusion processes.

    \item We benchmark our model on the unitary compilation problem, showcasing its capabilities across varying qubit counts and circuit depths. When comparing to existing approaches, our method excels at generating shorter circuits, although at the expense of lower accuracies, making it a promising tool for NISQ devices. We then present a simple search approach that boosts the accuracy of the method.  
    
    \item Leveraging the fast generation capabilities of our method, we generate large circuit datasets for specific unitaries and use these to reveal interesting structural patterns. For instance, our method is able to find the textbook circuit for the quantum Fourier transform.
\end{enumerate}

\section{Preliminaries} \label{sec:preliminaries}

\paragraph{Unitary compilation}
The unitary compilation problem is defined by two inputs: a complex-valued unitary matrix $\mathcal{U}$ of size $2^n \times 2^n$, where $n$ is the number of qubits, and a finite gate set, often related to the available physical operations of the quantum hardware onto which we aim to compile the unitary. Gates can be either fixed (e.g., the Hadamard or CNOT gates) or parameterized, in which case they are defined by a type and a parameter $\theta$ that can vary continuously (e.g., the X-rotation gate $R_x(\theta)$, with $\theta \in [0, 4\pi]$~\cite{nielsen2010quantum}). In the following, we will also refer to the former as discrete gates, and to the latter as both parametrized and continuous gates. We note that any discrete gate can be decomposed into continuous gates, but allowing the usage of both gate types can help get better and more interpretable circuits (see \cref{sec:discovery}). The output of the compilation problem is a quantum circuit, i.e., a sequence of quantum gates drawn from the specified gate set that implements $\mathcal{U}$ on quantum hardware. Given a circuit $C$, we define its distance to the target unitary $\mathcal{U}$ via its unitary representation $\mathcal{U}_C$, using the infidelity:
\begin{equation}
	\label{eq:infidelity}
	\mathcal{I} = 1-\frac{1}{4^n}|\mathrm{Tr}(\mathcal{U}^\dagger_C \mathcal{U})|^2,
\end{equation}
with $\mathcal{I}=0$ for $\mathcal{U}_C=\mathcal{U}$ up to a global phase, and $0\leq\mathcal{I}\leq 1$.
Unitary compilation is a fundamental task in quantum computation and has therefore received significant attention. While the Solovay-Kitaev theorem ensures \emph{efficient} compilation for single-qubit gates~\cite{dawson2005solovay}, no guarantees exist for multi-qubit systems. In this regime, a variety of heuristic, search-based, gradient-based, and machine learning methods have been explored. Approaches such as~\cite{davis2020towards, madden2022best, du2022quantum, grimsley2019adaptive} alternate between searching for a circuit structure and optimizing gate parameters until they reach a desired error threshold. These methods achieve high fidelity but at the expense of long runtimes and deep circuits. In recent years, machine learning has had a substantial impact on the field~\cite{krenn2023artificial}. For example, Ref.~\cite{he2023gnn} proposed the use of graph neural networks to improve the ansatz selection within the aforementioned iterative framework, while~\cite{ostaszewski2021reinforcement} achieves this via reinforcement learning (RL). In the latter, an agent is trained to sequentially place gates in a circuit and is rewarded when the resulting unitary closely approximates the target (see also ~\cite{bolens2021reinforcement, melnikov2018active, rietsch2024unitary, moro2021quantum, olle2024simultaneous,
preti2024hybrid}). 

Importantly, all previous methods consider cost functions or rewards that require the simulation of the proposed circuit, a classically hard computation which motivated the development of methods working natively on quantum computers~\cite{khatri2019quantum}. Another approach to circumvent this bottleneck is to train a generative model on a dataset of labeled circuits, as done for instance in Ref.~\cite{nakaji2024generative}. Although generating and labeling such a dataset is computationally expensive, once it is built, all further training is purely classical. In this direction, Ref.~\cite{furrutter2024quantum} proposed the use of DMs for quantum circuit synthesis, albeit demonstrated only on discrete gate sets and circuits with up to three qubits for unitary compilation.

\paragraph{Gaussian diffusion models}
Diffusion models (DM) are a class of generative models that aim to learn the underlying probability distribution $p_{\text{data}}(\vb{x})$ of a dataset. Starting from a clean data sample $\vb{x}$, the forward diffusion process defines a sequence of latent variables $\{\vb{z}_t\}_{t\in [0,1]}\coloneq \vb{z}_{0:1}$, where each $\vb{z}_{t}$ is a progressively noisier version of $\vb{x}$ starting at $t = 0$. The objective of the model is to recover the original sample from a noisy observation $\vb{z}_t$. In continuous domains, this process can be formalized either as a stochastic differential equation (SDE)~\cite{karras2022elucidating} or equivalently as a discrete-time Markov chain~\cite{ho2020denoising}. In the case of Gaussian diffusion, the transition probability from $\vb{z}_t$ given $\vb{x}$ is defined as
\begin{equation}\label{eq:gaussian_dm}
	q(\vb{z}_t|\vb{x}) = \mathcal{N}(\vb{z}_t; \sqrt{\bar{\alpha}_t}\vb{x}, \bar{\beta}_t\vb{I}),
\end{equation} 
where the components of $\vb{x}$ are independently corrupted and $\bar{\alpha}_t, \bar{\beta}_t \in [0,1]$ are the noise schedule coefficients, monotonically decreasing and increasing functions of $t$. A common choice is the variance-preserving model, where $\bar{\beta}_t = 1-\bar{\alpha}_t$. From here, one defines the signal-to-noise-ratio (SNR) as $\mathrm{SNR}(t)\coloneq {\bar{\alpha}_t}/\br{1-\bar{\alpha}_t}$. In this notation, the diffusion process maps samples with negligible noise at small times ($\lim_{t\rightarrow0}\mathrm{SNR}(t)\rightarrow\infty$) to fully noisy representations ($\mathrm{SNR}(t=1)=0$).

Diffusion models are trained by maximizing the evidence lower bound (ELBO) of the log-likelihood of $\vb{x}$ ~\cite{ho2020denoising}. Once the DM is trained, sampling is typically performed via ancestral sampling: starting from a fully complete noisy latent $\vb{z}_1\sim\mathcal{N}(\vb{0}, \vb{I})$, one follows the reverse transitions $p(\vb{z}_s|\vb{z}_t)$ for $0\leq s < t \leq 1$ until reaching a clean sample at $t=0$, as for instance described in Ref.~\cite{ho2020denoising}. 

\begin{figure*}
	\centering
	\includegraphics[width=0.95\textwidth]{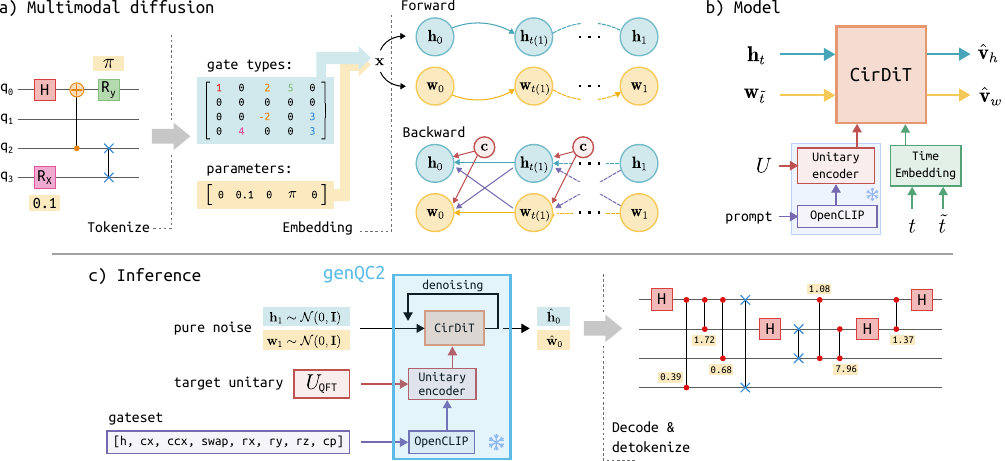}
	\caption{\textbf{Multimodal quantum circuit synthesis pipeline scheme.}
		\textbf{a)} An input circuit is first tokenized and then embedded into two separate modes, from which the forward and backward diffusion processes are defined.
		\textbf{b)} Schematic of the generative model.
		\textbf{c)} Inference overview, exemplary for the quantum Fourier transform. See \cref{sec:methods} for details.
	}
	\label{fig:DM_overview}
\end{figure*}

\paragraph{Multimodal diffusion models}
Multimodal diffusion models (DMs) address the generation of samples composed of multiple modes. Recent works explored embedding multimodal data into a joint latent representation and applying a single Gaussian diffusion process~\cite{chen2024diffusion, si2025tabrep}. Other approaches rather consider two identical Gaussian processes~\cite{bao2023one, ruan2023mm}. On the other hand, previous works have also explored generation via separated diffusion processes as, e.g., in Ref.~\cite{kotelnikov2023tabddpm} by combining Gaussian with multinomial diffusion or in Ref.~\cite{shi2024tabdiff} using a Gaussian and a masked diffusion process.

\section{Methods} \label{sec:methods}

\subsection{Multimodal diffusion process}
\label{sec:multidiff}
To perform simultaneous generation of discrete and parameterized gates, we represent each quantum circuit  $\vb{x}$ as a combination of two different modes: a discrete (categorical) mode $\vb{h}_0$  that encodes the gate types, and a continuous mode $\vb{w}_0$ that specifies the values of the parameterized gates. We then design the forward process of the DM as two independent diffusion processes, each acting on its respective data mode (\cref{fig:DM_overview}a). This separation allows us to construct mode-specific embeddings, tailored to the characteristics of each data type. In particular, we set the joint probability of the trajectories $\vb{h}_{0:1}$ and $\vb{w}_{0:1}$ of the forward process as the product of the independent joint distributions:
\begin{equation} \label{eq:forward_multimodial}
	q(\vb{h}_{0:1}, \vb{w}_{0:1} | \vb{x}) = q(\vb{h}_{0:1} | \vb{x})\; q(\vb{w}_{0:1} | \vb{x}),
\end{equation}
where the separate distributions are defined via the standard Gaussian diffusion (see \cref{eq:gaussian_dm}). Although the forward process in \cref{eq:gaussian_dm} is already component-independent, we further consider here that each diffusion process has a different noise schedule, namely $\bar{\alpha}_{t}^h$ and $\bar{\alpha}_{t}^w$, with the corresponding signal-to-noise ratios $\mathrm{SNR}_h(t)\coloneq\bar{\alpha}_t^{h}/(1-\bar{\alpha}_t^{h})$ and $\mathrm{SNR}_w(t)\coloneq\bar{\alpha}_t^{w}/(1-\bar{\alpha}_t^{w})$.  

This mode separation is motivated by the observation that both the choice of diffusion schedule and the design of data embeddings play a crucial role in shaping the behavior of the diffusion process. In particular, as categorical data are sparsely distributed in the latent space, it requires tuned noise schedules that gradually mix the different classes before reaching full noise. 
In this sense, accurately selecting both the noise schedule and its weighting in the loss has a significant influence on the training efficiency of DMs, as often studied~\cite{nichol2021improved, kingma2023understanding, hang2023efficient, hang2024improved, hoogeboom2024simpler}. In this direction, we show in \cref{sec:embedding} and \cref{sec:learned_sched} that to effectively model discrete data in a Gaussian diffusion model, it is crucial to choose a suitable noise schedule, and present a learning procedure to appropriately match the token embeddings.

\begin{figure*}
	\centering
    \includegraphics[width=0.95\textwidth]{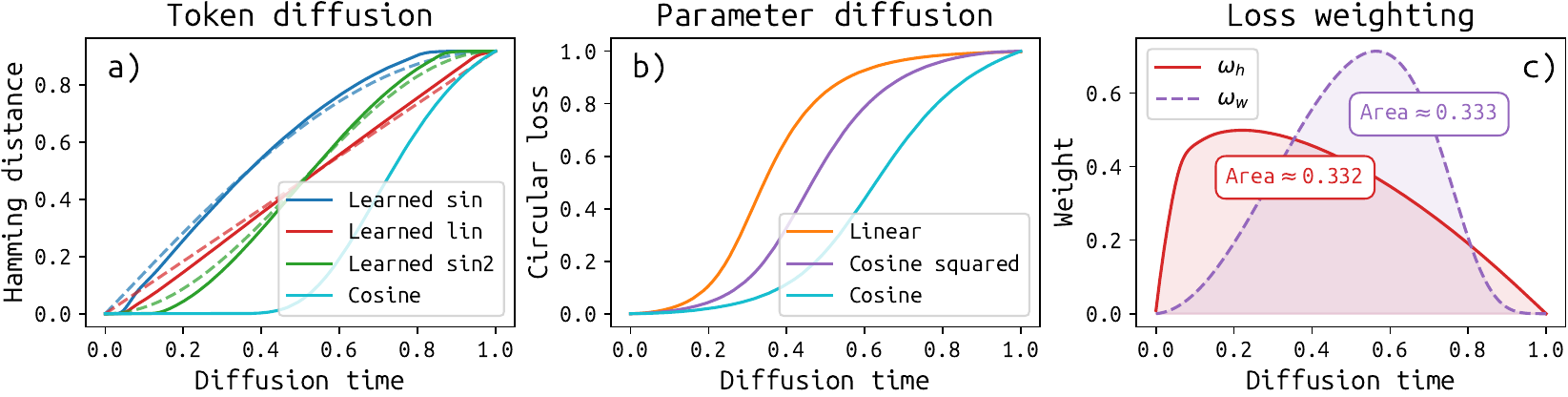}
	\caption{\textbf{Noise schedules and loss weighting.}
		\textbf{a)} Averaged Hamming distance between an initial token $\vb{h}_0$ and the decoded embedding of $\vb{h}_t$ over the diffusion time $t$. Dashed lines represent target schedules and solid lines represent the learned ones (see \cref{sec:learned_sched}).	
		\textbf{b)} Circular loss between initial parameter $\vb{\lambda}$ and the decoded parameter $\hat{\lambda}_t$ over the diffusion time $t$ for different noise schedules.
		\textbf{c)} Loss weighting for the discrete ($\omega_h(t)$) and continuous ($\omega_w(t)$) modes used in \cref{eq:loss} over the diffusion time, chosen such that their total areas roughly match.
		}
	\label{fig:DM_schedule}
\end{figure*}

\paragraph{Loss}
Using the velocity parametrization picture~\cite{salimansprogressive}, the model is trained by minimizing the loss function (see details in \cref{sec:app_multimode_details})
\begin{multline}\label{eq:loss}
	\expecOp_{
		(\vb{x}, \vb{c})\sim p_\text{data},\,
		(t,\tilde{t})\sim\mathcal{U}\br{0,1},\,
		(\vb{h}_t, \vb{w}_{\tilde{t}})\sim q(\vb{h}_t, \vb{w}_{\tilde{t}}|\vb{x})
	}\\ \Big[
	\omega_h(t)\norm{\vb{v}_t^{h} - \vb{v}_{\theta}^{h}}_2^2 +
	\omega_w(\tilde{t})\norm{\vb{v}_{\tilde{t}}^{w} - \vb{v}_{\theta}^{w}}_2^2
	\Big],
\end{multline}
where the labeled pairs $\br{\vb{x}, \vb{c}}$ are samples from the dataset distribution $p_\text{data}$. The two diffusion times, $t$ and $\tilde{t}$, are independently drawn from a uniform distribution. The noisy latents $\vb{h}_t$ and $\vb{w}_{\tilde{t}}$ are sampled from the independent Gaussian forward processes (see \cref{eq:forward_multimodial}), each being sampled via \cref{eq:gaussian_dm}. The velocity parametrization replaces the predictor $\hat{\vb{x}}_\theta$ by $\hat{\vb{v}}_{\theta}$, which outputs jointly $[\vb{v}^h_{\theta}, \vb{v}^w_{\theta}]=\hat{\vb{v}}_{\theta}(\vb{h}_{t}, \vb{w}_{\tilde{t}}, t, \tilde{t}, \vb{c})$ (see \cref{fig:DM_overview}b). The corresponding targets are then set to $\vb{v}_t^{h}=\sqrt{\bar{\alpha}_t^{h}}\,\vb*{\epsilon}^{h}-\sqrt{1-\bar{\alpha}_t^{h}}\,\vb{h}_0$ and $\vb{v}_{\tilde{t}}^{w}=\sqrt{\bar{\alpha}_{\tilde{t}}^{w}}\,\vb*{\epsilon}^{w}-\sqrt{1-\bar{\alpha}_{\tilde{t}}^{w}}\,\vb{w}_0$. This choice of the target velocities ensures that the model outputs have unit variance, i.e., $\varOp[\vb{v}_t^{h}]=\varOp[\vb{v}_{\tilde{t}}^{w}]=1$, assuming that the embeddings are normalized such that $\varOp[\vb{h}_0]=\varOp[\vb{w}_0]=1$. 

For the weighting terms $\omega_h(t)$ and $\omega_w(\tilde{t})$, we use the sigmoid method introduced in~\cite{hoogeboom2024simpler}, which relates the weight in the denoising prediction view $\hat{\vb{x}}_{\theta}$ to the SNR, given by $\omega_x(t)=\mathrm{sigmoid}\bre{\log(\mathrm{SNR}(t))}$. For velocity prediction, this relation has to be translated with $\omega_x(t)=(\mathrm{SNR}(t)+1)\,\omega_v(t)$ (see \cref{sec:app_v_weight}). Since we only use velocity prediction, we omit the $v$ subscript from here on, and incorporate the corresponding factor $(\mathrm{SNR}_{i}(t)+1)^{-1}=(1-\bar{\alpha}_{t}^{i})$ into  $\omega_i(t)$ for $i\in\{h, w\}$ when defining the weights in \cref{eq:loss}. In \cref{sec:embedding} and \cref{sec:learned_sched}, we generalize the log-SNR to both discrete and continuous modes, providing definitions for $\omega_h(t)$ and $\omega_w(t)$ accordingly.

\paragraph{Unitary condition and inference}
In this work, the model's conditioning $\vb{c}$ consists of two parts: the unitary $\mathcal{U}$ to be synthesized and a text prompt specifying the available gates (see \cref{fig:DM_overview}b). We first create the embeddings of the text prompts with a pretrained \textit{OpenCLIP}~\cite{ilharcoOpenCLIP}. Second, we train a custom \textit{UnitaryCLIP} which contrastively learns to encode a quantum circuit and a unitary together with the gate set embedding into a joint latent space (see \cref{sec:app_unitary_clip}). After that, the unitary encoder model is frozen and the latent embeddings arising from given unitaries and prompts are used as conditioning of the DM. For inference, we start with pure noisy embeddings $\vb{h}_1,\vb{w}_1\sim\mathcal{N}(0,\vb{I})$, which get iteratively denoised and then decoded back to a circuit (see \cref{fig:DM_overview}c).

\subsection{Circuit encoding and embedding}\label{sec:embedding}
In this section, we describe the method used to transform a quantum circuit $\vb{x}$ into a suitable embedding. For each mode, we consider fixed deterministic encoders $q(\vb{h}_0|\vb{x})$ and $q(\vb{w}_0|\vb{x})$. The decoder $p(\vb{x}| \vb{h}_{0}, \vb{w}_{0})$ is likewise fixed. Following Ref.~\cite{furrutter2024quantum}, the gates of a given circuit are first tokenized (see \cref{fig:DM_overview}a). Then, each gate is represented by a tuple $(k, \lambda)$, where $k$ is the token and $\lambda\in\bre{-1,1}$ the normalized continuous parameter, equal to zero for discrete gates.

\paragraph{Token embedding} 
Token embeddings are typically implemented as look-up tables containing $d_h$-dimensional entries $\vb{h}_0^{(i)}\in\reals^{d_h}$, where $i  = {0, \dots, N-1}$ indexes the $N$ different classes. During decoding, a given embedding is mapped to the closest token in the table based on a distance metric. To ensure that all embeddings are equidistant and undergo uniform mixing throughout the diffusion process, we construct them as an orthogonal basis of $\reals^{d_h}$ (more in \cref{sec:app_discr_emb}). This orthogonality not only balances the embedding space but also enables us to leverage the duality between uniform discrete-state diffusion and continuous-state Gaussian diffusion~\cite{sahoo2025diffusionduality}. 

To study the impact of the noise schedule $\bar{\alpha}_t^{h}$ on these embeddings, we plot in \cref{fig:DM_schedule}a the averaged Hamming distance between the decoded embedding at each step and the original token. We observe that standard schedules, such as the cosine schedule~\cite{nichol2021improved}, fail to sufficiently mix the embeddings, causing the decoded token to remain unchanged over extended times of the diffusion process. This behavior leads to inefficient training, as little signal is provided for learning. Ideally, the noise schedule should induce gradual and consistent mixing across time. To this end, in \cref{sec:learned_sched} we introduce a method for learning a noise schedule tailored to the desired token embeddings' mixing.

\paragraph{Parameter embedding}
The continuous parameters $\lambda$ are embedded into a two-dimensional plane following $\vb{w}_0(\lambda) = \cos(\lambda\pi) \vb{v}_1 + \sin(\lambda\pi) \vb{v}_2$, where $\vb{v}_1, \vb{v}_2\in\reals^{d_w}$. This accounts for the periodic rotation angles of the gates. An encoded noisy $\vb{w}_t(\lambda)$ can be decoded back to a parameter $\hat{\lambda}_t$ by using the $\mathrm{arctanh2}$ (see \cref{sec:app_cont_emb}). As done above, we study the effect of different noise schedules on such an embedding. We analyze in this case the $\mathrm{CircularLoss}(t) \coloneq 1 - \mathrm{cos}((\lambda-\hat{\lambda}_t)\pi)$ (\cref{fig:DM_schedule}b). As expected, the choice of noise schedule influences the rate at which the decoded parameter $\hat{\lambda}_t$ approaches the uniform distribution. Following the previous definition, the SNR associated with this embedding can be estimated as
$
\mathrm{SNR}_\lambda(t) \coloneq \mathrm{SNR}_w(t) \cdot \pi^2 d_w.
$
Following this, we set $\omega_w(t)\coloneq (1-\bar{\alpha}_{t}^{w}) \cdot\mathrm{sigmoid}[\log(\mathrm{SNR}_\lambda(t))]$ in \cref{eq:loss} (see full derivation in \cref{sec:app_cont_snr}). 

\subsection{Learned noise schedule for discrete tokens} \label{sec:learned_sched}

Following the observations of the previous section, we construct here a noise schedule  $\bar{\alpha}_t^h$  that gradually mixes the input gate tokens, minimizing the time of trivial denoising. Using the duality between uniform discrete-state diffusion and continuous-state Gaussian diffusion~\cite{sahoo2025diffusionduality}, we write the probability of finding a decoded, one-hot encoded token $\vb{k}_t\in\reals^N$ for a discrete schedule $a_t$ as
\begin{equation}\label{eq:cat_DM}
	p(\vb{k}_t|\vb{k}) = \mathrm{Cat}(\vb{k}_t; a_t \vb{k} + (1-a_t)\vb{I}/N).
\end{equation}
From this, we define the SNR of the discrete mode as the fraction of the original amplitude to the one of the uniform distribution: $\mathrm{SNR}_{\text{discrete}}(t) \coloneq {a_t}/({1 - a_t})$. In previous approaches~\cite{sahoo2025diffusionduality}, one first specifies the Gaussian diffusion schedule $\bar{\alpha}_t^h$, to then obtain $a_t$. Here, we take the reverse approach:  we begin by specifying the desired level of mixing among discrete tokens, and then infer the corresponding $\bar{\alpha}_t^h$.
 
To this end, we define an analogue to the average Hamming distance $f(t)$, namely the probability $p_\text{flip}(t)$ of a token initially belonging to class $i$ being decoded as any other class $j\neq i$ at time $t$:
\begin{multline} \label{eq:p_flip}
	p_\text{flip}(t) = 1 - \expecOp_{\vb{h}_t^{(i)}\sim q(\vb{h}_t^{(i)}|\vb{h}_0^{(i)})} \\ \Big[
	{
		\mathrm{softmax}_j\br{\frac{1}{\tau}\braket{\vb{h}_0^{(j)}}{\vb{h}_t^{(i)}}}
	}\Big]_i,
\end{multline}
where $\tau>0$ is a temperature. Analogous to the average Hamming distance, this probability is upper bounded by $1-1/N$, i.e., when $\vb{h}_t^{(i)}$ is sampled from the uniform distribution (see $t=1$ at \cref{fig:DM_schedule}a).

We then use this definition to learn the appropriate coefficients $\bar{\alpha}_t^h$ in $q(\vb{h}_t^{(i)}|\vb{h}_0^{(i)})$ (see \cref{eq:gaussian_dm}) for a desired Hamming distance $f_\text{target}(t)$. This is achieved by minimizing the mean-squared error loss  $\expecOp[\norm{p_\text{flip}(t)-f_\text{target}(t)}^2]$ w.r.t. $\bar{\alpha}_t^h$. In \cref{fig:DM_schedule}a we show a few examples of desired $f_\text{target}(t)$ (dashed lines) and their corresponding optimized Hamming distance profiles (solid lines). This training  occurs prior to the training of the diffusion model itself, and offers the advantage of directly enforcing the desired token mixing behavior, bypassing the need for manual tuning of known schedules. Next, to determine the appropriate weight $\omega_h(t)$ in \cref{eq:loss}, we combine \cref{eq:cat_DM} and \cref{eq:p_flip} to find that we get the relation $a_t = 1 - p_\text{flip}(t) / p_\text{flip}(1)$ (see \cref{sec:app_learned_schedule}). From this, we derive
$
	\mathrm{SNR}_{\text{discrete}}(t)  \approx (f_\text{target}(1) - f_\text{target}(t))/f_\text{target}(t),
$
which we then use to define $\omega_h(t)\coloneq (1-\bar{\alpha}_{t}^{h}) \cdot\mathrm{sigmoid}[\log(\mathrm{SNR}_{\text{discrete}}(t))]$. 

Finally, to balance the two loss terms in \cref{eq:loss}, we compute the area under each weighting curve $\omega_h(t)$ and $\omega_w(t)$ (\cref{fig:DM_schedule}c). We then select noise schedules for the discrete and continuous modes such that their corresponding areas are approximately equal, thereby implicitly balancing the two expectations without requiring an additional weighting factor.

\subsection{Gate-Pair tokenization}
\label{sec:gpe}
In natural language processing, \textit{Byte-Pair Encoding} (BPE)~\cite{gage1994new} is a widely used technique for subword tokenization. Since our circuits are already tokenized, it is natural to extend BPE to the domain of quantum circuits. To this end, we generalize the notion of byte pairs to pairs of consecutive quantum gates that are sequential—i.e., those acting on at least one common qubit. Once candidate gate pairs are identified, we normalize their qubit connections to account for permutations, ensuring that we capture generalizable patterns of higher-order gates rather than qubit-specific configurations. We then count the frequency of each pair, select the most frequent one, assign it a new token, and replace all occurrences in the current dataset, repeating the process until reaching a predefined minimum frequency threshold. In \cref{sec:discovery}, we demonstrate how this \textit{Gate-Pair Encoding} (GPE) scheme can be employed to automatically extract reusable substructures (gadgets) from generated circuits corresponding to specific unitaries.

\section{Experiments}\label{sec:experiments}

\begin{figure*}
	\centering
	\includegraphics[width=0.95\textwidth]{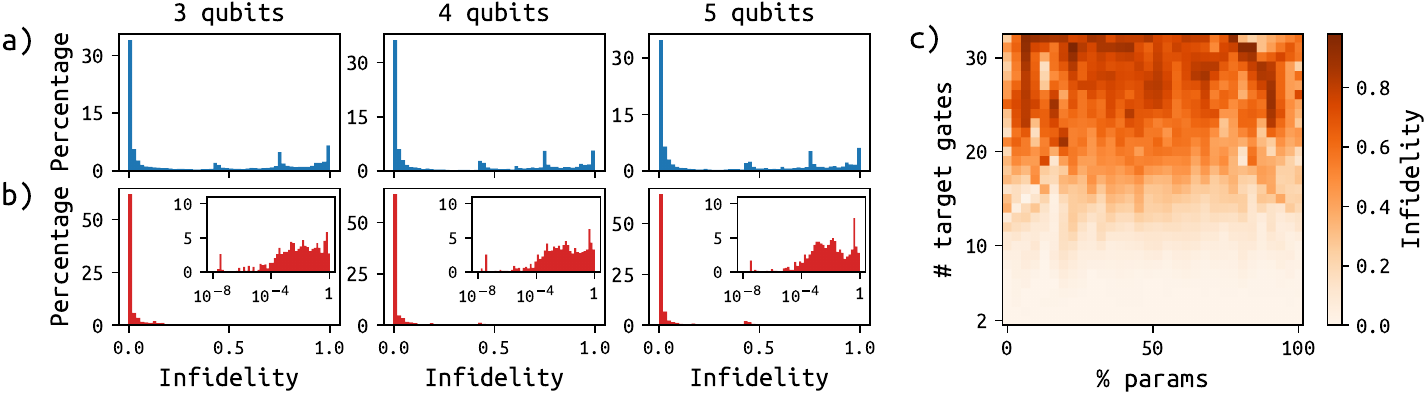} 
	\caption{\textbf{Synthesis of random unitaries.}
		\textbf{a)} Histogram of infidelities for 1024 unitaries and 128 circuits sampled per unitary (up to 16 gates).
		\textbf{b)} Histogram of minimum infidelity for each of the 1024 unitaries in a). The insets showcase the same plot but in logarithmic scale.
		\textbf{c)} Best infidelity over 128 circuits for target unitaries of varying gate count and percentage of parameterized gates, averaged for 3 to 5 qubits.
	}
	\label{fig:compile_rnd}
\end{figure*}

\begin{table*}
	\caption{\textbf{Synthesis methods comparison.} Reported values are averages over random 480 unitaries (from 2 to 16 gates) and 128 circuits sampled for each unitary. Values in parenthesis are the corresponding standard deviation, showing the variability of the metrics.(*)~Some sample restrictions apply, as detailed in \cref{sec:comparison}. (**)~Not reported due to lower sample count.
    }
	\label{tab:compiler_comparison}
	\centering
	\begin{tabular}{
			p{2.15cm}
			c
			>{\raggedleft\arraybackslash}p{3.6cm}
			>{\raggedleft\arraybackslash}p{1.6cm}
            >{\raggedleft\arraybackslash}p{2cm}
			>{\raggedleft\arraybackslash}p{2.0cm}
		}
		\toprule
        Method & \# qubits & Avg. minimum\par infidelity & Avg. gate count & Avg. runtime per sample (seconds) & Avg. distinct circuits per unitary \\ 

        \midrule	
		genQC2 (ours)		& 3 & $0.09\;(0.20)$ & $8\;(4)$ & $0.09\;(0.00)$ & $78\;(45)$ \\ 
		~ 					& 4 & $0.10\;(0.20)$ & $8\;(4)$ & $0.09\;(0.00)$ & $74\;(49)$ \\ 
		~ 					& 5 & $0.09\;(0.17)$ & $8\;(4)$ & $0.09\;(0.00)$ & $71\;(50)$ \\ 
        
        \midrule		
		+ ansatz fixes & 3 & $0.008\;(0.037)$ & $ 9\;(5)$ & $190\;(390)$ & NA \\ 
		  (\cref{sec:gen_ansatz}) & 4 & $0.015\;(0.062)$ & $ 9\;(5)$ & $330\;(570)$ & NA \\ 
		  ~ & 5 & $0.015\;(0.071)$ & $ 9\;(5)$ & $480\;(880)$ & NA \\ 

        \midrule   
		QSD 	& 3 & $0.4\cdot 10^{-15}\;(2.1\cdot 10^{-15})$ & $68\;(15)$ & $0.04\;(0.01)$ & $1\;(0)$ \\ 
		(Refs.~\cite{shende2006qsd, krol2024beyond}) & 4 & $0.2\cdot 10^{-10}\;(4.4\cdot 10^{-10})$ & $348\;(24)$ & $0.16\;(0.01)$ & $1\;(0)$ \\ 
		~ 					& 5 & $0.1\cdot 10^{-10}\;(1.2\cdot 10^{-10})$ & $1558\;(46)$ & $0.75\;(0.03)$ & $1\;(0)$ \\ 

        \midrule  
		AQC*                         & 3 & $1.0\cdot 10^{-9}\;(3.8\cdot 10^{-9})$ & $79\;(0)$   & $0.3\;(0.2)$  & $21\;(12)$ \\ 
		(Ref.~\cite{madden2022best}) & 4 & $4.3\cdot 10^{-5}\;(5.4\cdot 10^{-5})$ & $317\;(0)$  & $5.2\;(0.1)$  & ~$^{**}$ \\ 
		~ 	                         & 5 & $1.0\cdot 10^{-4}\;(0.3\cdot 10^{-4})$ & $1275\;(0)$ & $27.7\;(0.1)$ & ~$^{**}$ \\ 

        \midrule    
		LEAP*   & 3 & $0.4\cdot 10^{-3}\;(1.9\cdot 10^{-3})$ & $61\;(26)$ & $0.8\;(0.6)$ & $41\;(39)$ \\ 
		(Ref.~\cite{smith2023LEAP}) & 4 & $1.2\cdot 10^{-3}\;(3.2\cdot 10^{-3})$ & $85\;(50)$ &  $40\;(250)$ & ~$^{**}$ \\ 
		~ 				& 5 & $2.7\cdot 10^{-3}\;(5.3\cdot 10^{-3})$ & $92\;(57)$ & $170\;(500)$ & ~$^{**}$ \\ 

		\bottomrule
	\end{tabular}
\end{table*}

\subsection{Experimental Setup}\label{sec:experiments_setup}
We present here the main information on the training of the model discussed in previous sections. Further details can be found in \cref{sec:app_training_details}. 

\paragraph{Training dataset}
Using CUDA-Q~\cite{cudaq}, we generate a training dataset of random 3 to 5 qubit circuits. To that end, we uniformly sample 4 to 32 gates from the gate set $\brek{\mathrm{h},\mathrm{cx},\mathrm{ccx},\mathrm{swap},\mathrm{rx},\mathrm{ry},\mathrm{rz},\mathrm{cp}}$. The continuous parameters of the parameterized gates are also sampled uniformly on their support. The model used throughout this work is trained on a dataset of 63 million unitary-circuit pairs. 

\paragraph{Models, training and inference}
The model presented here, named Circuit-Diffusion-Transformer (CirDiT), is based on the diffusion transformer architecture~\cite{peebles2023scalable}, and contains 150 million parameters, for details see \cref{sec:app_model_arch}. Training is performed on 16 NVIDIA A100 GPUs for a preset number of $\sim$800k update steps, with an effective batch size of 2048. The effective training time is roughly 700 single GPU hours. We use the \textit{Adam} optimizer~\cite{kingma2014adam} together with a one-cycle learning rate strategy~\cite{smith2019super}. We use a learned linear noise schedule for the discrete mode (see \cref{sec:learned_sched}) and a fixed schedule for the continuous mode. Once trained, we sample from the model using the CFG++~\cite{chung2024cfgPP} variant of the DPM++2M~\cite{lu2022dpm} solver for 40 time steps.

\subsection{Benchmark on random unitaries}
\label{sec:benchmark}

We first benchmark the model by synthesizing circuits from a test set of unitaries mimicking the properties of the training dataset (see \cref{sec:app_testset} for details). We then compute the infidelity, \cref{eq:infidelity}, between the unitary representation of the output circuit and the target one. To exclude overfitting, we report infidelities for training set unitaries in \cref{sec:overfitting}.

We present in \cref{fig:compile_rnd}a and b the distribution of infidelities for all generated circuits and that of the circuits with the lowest infidelity, respectively, using target unitaries of 3 to 5 qubit circuits consisting of up to 16 gates, demonstrating that the model successfully compiles these unitaries with low infidelity. We observe characteristic peaks in the infidelity distribution around $\mathcal{I} = 0.4$, $0.8$, and $1$, which are attributed to the misplacement of one or two discrete gates, rather than errors in the continuous parameters of parameterized gates (see  \cref{sec:app_circuit_corruption}). Moreover, we also observe that the model’s accuracy remains stable with increasing qubit count. However, we observe a significant drop in performance as gate count increases.

To further investigate these trends, we show in \cref{fig:compile_rnd}c the infidelity as a function of the gate count and the percentage of parameterized gates. As mentioned, we observe a strong dependence on the number of gates: deeper circuits are harder to compile accurately. In contrast, we find no significant correlation between accuracy and the percentage of parameterized gates, indicating that the model handles both discrete and continuous gates with comparable effectiveness.

\subsubsection{Comparison to existing methods}
\label{sec:comparison}

We now want to compare our approach, named genQC2, to other methods for unitary synthesis. We provide a comparison to the previous diffusion-based method, genQC1~\cite{furrutter2024quantum}, in \cref{sec:app_genqc_1vs2}. We note that, while reinforcement learning (RL) has achieved notable success in certain quantum circuit synthesis tasks~\cite{moro2021quantum, rietsch2024unitary, kremer2025optimizing}, to the best of our knowledge there is no method comparable to the one proposed here. First, these approaches are limited to discrete gates, which significantly constrains their expressive power. Second, they typically focus either on training for a single target unitary~\cite{preti2024hybrid} or for a fixed qubit count~\cite{rietsch2024unitary}.

Hence, we restrict the benchmark to three state-of-the-art synthesis methods: an exact approach based on the Quantum Shannon Decomposition (QSD)~\cite{shende2006qsd, krol2024beyond}, and two approximate approaches, the Approximate Quantum Compiler (AQC)~\cite{madden2022best} and the search-based approach LEAP~\cite{smith2023LEAP, bqskit}. We present in \cref{tab:compiler_comparison} averaged results over 480 unitaries, drawn randomly from test-set circuits from 2 up to 16 gates. The values in parenthesis are the corresponding standard deviations to show the variability of the metrics. As the metrics are strictly non-negative and the distributions heavily skewed (see \cref{fig:compile_rnd}b) the standard deviations cannot be interpreted as errors. We restrict all compilers to the same gate set as our model, i.e. $\brek{\mathrm{h},\mathrm{cx},\mathrm{ccx},\mathrm{swap},\mathrm{rx},\mathrm{ry},\mathrm{rz},\mathrm{cp}}$, and leave the number of allowed gates unlimited (see \cref{tab:compiler_comparison_restricted} for a comparison with limited gate count). We then sample 128 circuits for each unitary. However, as the runtime of AQC is relatively high, we restrict this method to 32 circuits per unitary for 4 qubits and 8 for 5 qubits. This is also the case for LEAP, where we restrict to 8 circuits per 4-qubit unitary and one for 5 qubits. Further, we set for LEAP a timeout of one hour per 5-qubit unitary, observing 28 ($6\%$) of the test unitaries did not finish within this time.

\begin{figure}
	\centering
	\includegraphics[width=0.78\columnwidth]{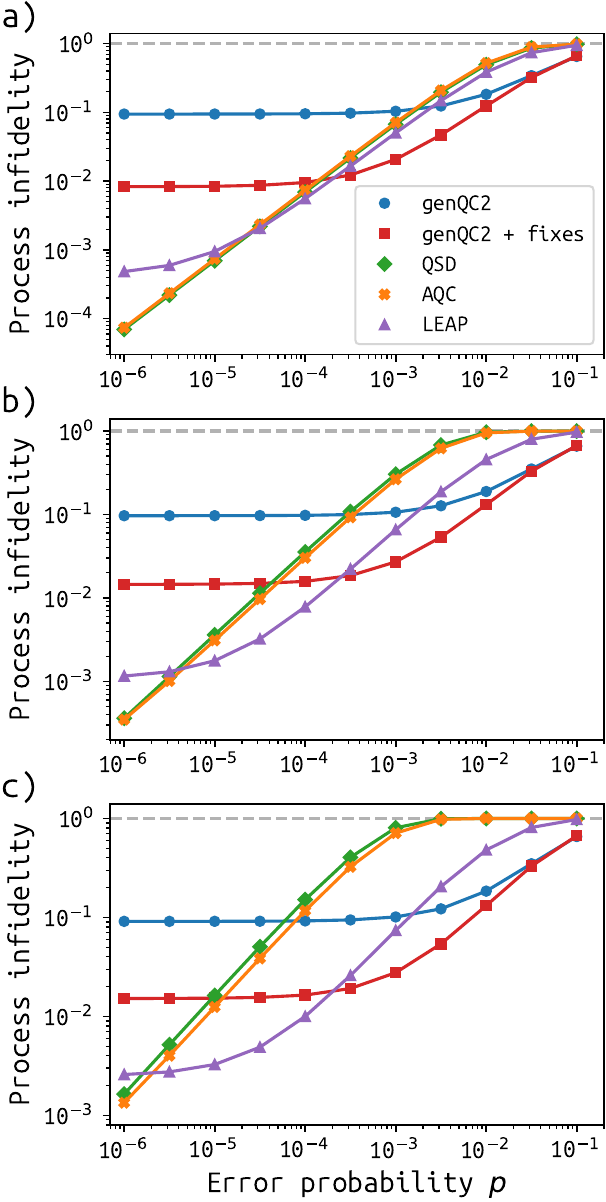}
	\caption{\textbf{Process infidelity under noise.} 
		Average minimum process infidelity of generated circuits under noisy simulation, depending on the value of the error probability $p$ (see \cref{sec:app_noisy_sim}). Plotted are infidelities for: \textbf{a)} 3, \textbf{b)} 4, and \textbf{c)} 5 qubits. The circuits are sampled the same way as in \cref{tab:compiler_comparison}.
	}
	\label{fig:noisy_simulation}
\end{figure}

We begin by noting that, as expected, the other synthesizers achieve significantly lower accuracies than genQC2, but at the cost of longer runtimes and larger circuits. In particular, the QSD and AQC compilers require between one and three orders of magnitude more gates than the target unitaries, which contain at most 16 gates. The search-based compiler generates shorter circuits than QSD and AQC but remains one to two orders of magnitude above the original depth, highlighting the inability of the compilers to produce optimized solutions. In contrast, genQC2 generates much shorter circuits, accurately matching the depth of the circuit the target unitary was generated from. Importantly, for practical applications on NISQ devices, it is crucial to minimize the gate count, as noise is injected for each applied gate. We analyze this in \cref{fig:noisy_simulation}, where we simulate the synthesized circuits under a depolarizing noise channel applied to every gate with an error probability $p$ (see \cref{sec:app_noisy_sim} for details). The reported process infidelities are calculated for the same circuits as in \cref{tab:compiler_comparison}, hence the infidelities in the low noise case $p\ll1$ are of the same order as those presented there. Notably, we clearly observe a crossover between the methods, making the circuits of genQC2 the superior choice when dealing with typical noise regimes. Further, as the gate count of the other methods increases significantly with the number of qubits, the crossing points shift to smaller noise levels (error probabilities) with more qubits.

\begin{figure*}
	\centering
	\includegraphics[width=0.95\textwidth]{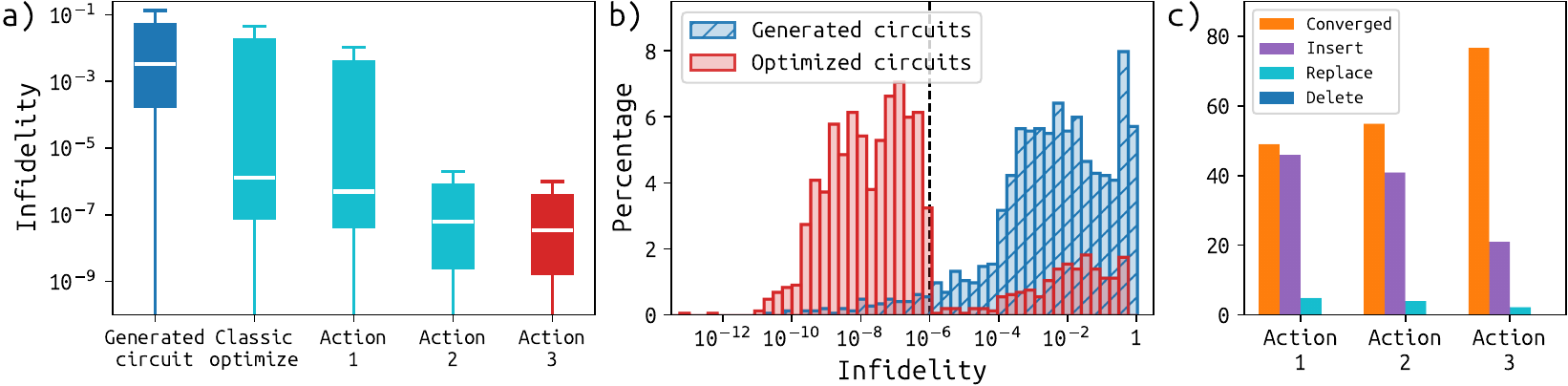}
	\caption{\textbf{Improving generative ansatz.} 
		Results after a tree-search with generative ansatz circuits as starting points.
		\textbf{a)} Boxplots of the infidelities of the generated circuits after performing some actions. White lines correspond to the medians.
		\textbf{b)} Histogram of infidelities before and after the improvements are applied.
        \textbf{c)} Distribution of taken actions at each tree depth. The converged action refers to nodes that have reached an infidelity threshold $\epsilon$ (vertical dashed line in panel b) and are not expanded anymore. 
	}
	\label{fig:ansatz_fixing}
\end{figure*}

Next, we compare the generation times. While runtime scaling can be meaningfully compared within each method in \cref{tab:compiler_comparison}, direct comparison of absolute runtimes across compilers is less straightforward, as further optimizations may exist for each implementation. Nevertheless, a clear trend emerges: although genQC2 requires a long training phase (see \cref{sec:experiments_setup}), once trained, sampling from it is extremely fast and does not depend on the number of qubits. The QSD compiler achieves comparable runtimes, though its execution time clearly increases with qubit count. In contrast, AQC and LEAP exhibit significantly longer runtimes. As noted earlier, we limited LEAP's maximum runtime to one hour, resulting in 28 ($6\%$) of the test unitaries not completing within this time and reducing the reported average runtime.

Finally, we examine the expressiveness of each method by tracking the number of distinct circuits produced per target unitary. As shown, a key advantage of our approach is its ability to generate multiple unique circuits for the same unitary, a feature we further exploit in \cref{sec:discovery}. While the QSD compiler is deterministic, yielding identical circuits across all 128 queries per unitary, genQC2 produces on average 74 distinct circuits per unitary. AQC and LEAP can also generate different circuits, though at a lower rate than genQC2.

\subsubsection{Improving generative ansatz}
\label{sec:gen_ansatz}

\begin{figure}
	\centering
	\includegraphics[width=0.9\columnwidth]{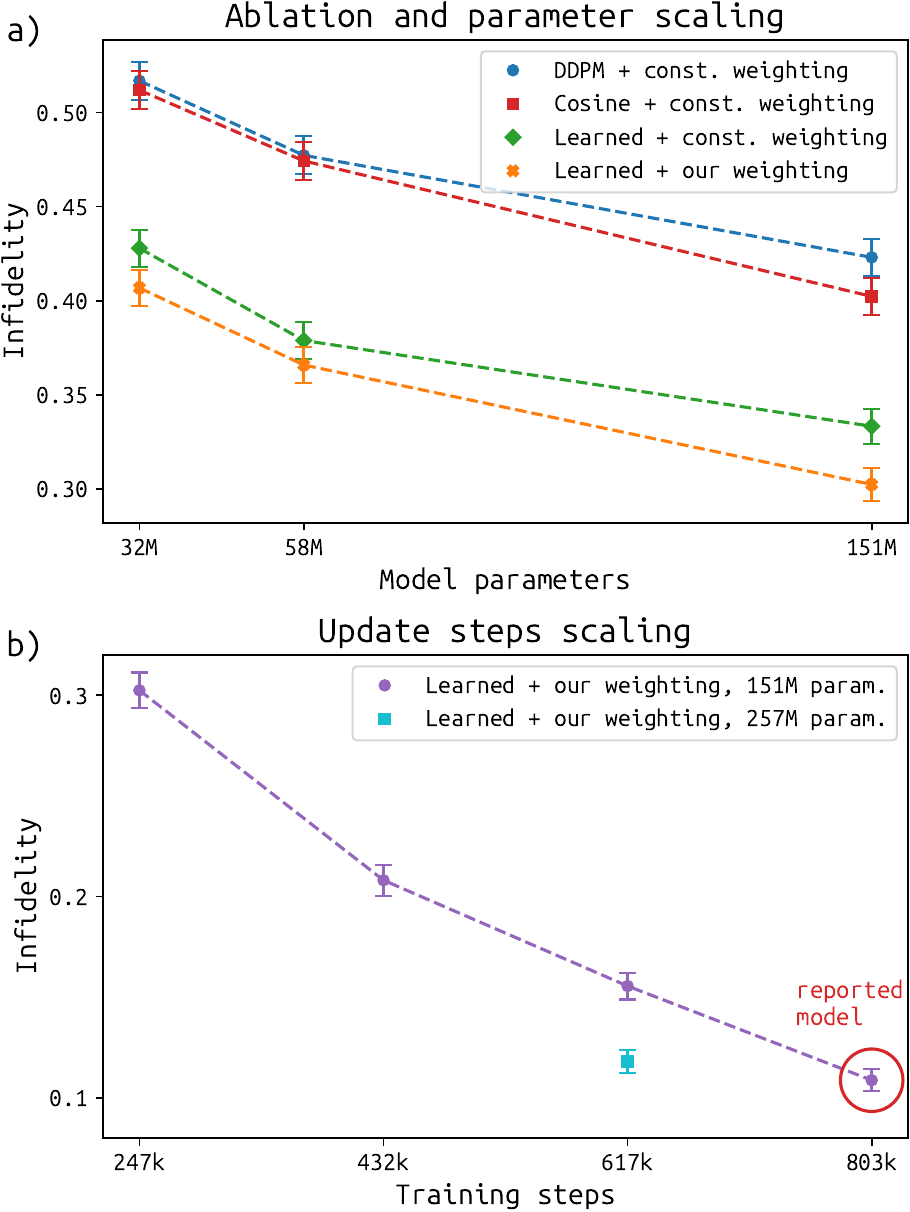} 
	\caption{\textbf{Ablation study and model scaling.}
		We sample for 3 to 5 qubits each 480 unitaries with 2 to 16 gates uniformly. For every unitary we generate 64 circuits and record the minimum infidelity. We show the averaged minimum infidelities for:
		\textbf{a)} different discrete noise schedulers and our adjusted weighting for different model  sizes;
		\textbf{b)} a different number of total training steps. Shown are different models, not a recording of a single model during training, as the learning rate schedule is depending on the total number of steps (see \cref{sec:app_training_details}).
		Errorbars are the errors of the means.
	}
	\label{fig:model_ablation_scale}
\end{figure}

\begin{figure*}[ht!]
    \centering
    \includegraphics[width=0.88\textwidth]{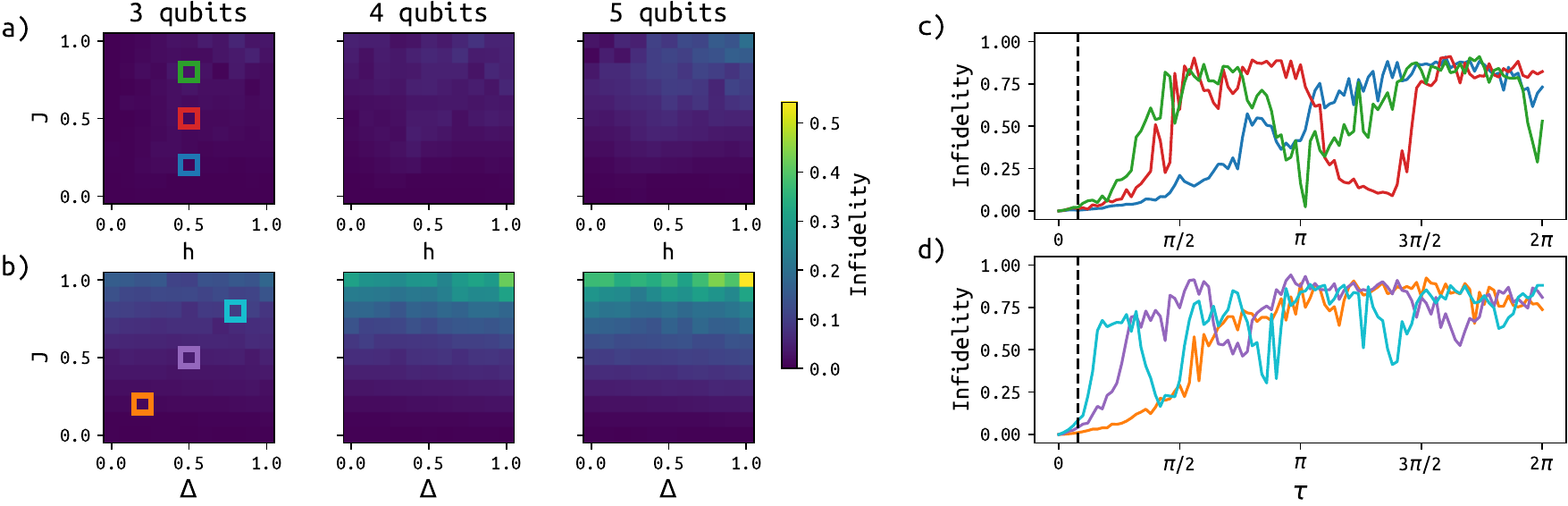} 
    \caption{\textbf{Hamiltonian evolution.} 
        \textbf{a, b)} Minimum infidelities over 128 circuits for different parameters of the Ising  and XXZ Hamiltonian, respectively, at $\tau=0.25$.
        \textbf{c, d)} Minimum infidelities over 128 circuits for different evolution times $\tau$ for both Hamiltonians with 3 qubits and various parameters, color matching the highlighted points in a and b panels. The vertical dashed is placed at $\tau=0.25$.
   	}
    \label{fig:hamiltonians}
\end{figure*}

While investigating the infidelity distribution of the method over different unitaries (\cref{fig:compile_rnd}a), we observed characteristic peaks which we attributed to one or two displaced discrete gates (see \cref{sec:app_circuit_corruption}). This suggests that the generated circuits, while partially incorrect, may be close to the target one. To test this, we take the generated circuits as ansätze to perform a shallow tree-search over possible gate additions, deletions or replacements (details in \cref{sec:app_tree_search}). If these ansatz circuits are indeed close to the exact solutions, a few actions should be able to lower the infidelity significantly. 

Given a generated circuit, we first perform a gradient-based optimization of the continuous angles, followed by a greedy top-$k$ expansion policy up to 3 actions (see \cref{sec:app_tree_search}). We stop expanding nodes which have reached an infidelity threshold $\epsilon = 10^{-6}$. Finally, we record the best trajectories and store the actions taken and the resulting infidelities. We note that recent works explore more efficient tree-search methods~\cite{lipardi2025quantum, valcarce2025unitary}, which can be applied to our generative ansätze, potentially leading to better solutions and faster synthesis, although their application falls out of the scope of this paper.

We present in \cref{fig:ansatz_fixing}a the infidelity distribution after each applied action, showing a final decrease over an order of magnitude w.r.t. the initial ansatz circuits. Notably, we observe that many ansätze can be improved by a large margin, as shown in the infidelity distribution before and after the improvements (see \cref{fig:ansatz_fixing}b). This is especially valuable as the model seeks exact solutions and not approximations, which significantly reduces the required number of gates compared to other compilers (see \cref{tab:compiler_comparison}). In \cref{fig:ansatz_fixing}c, we show the action distributions which led to the best circuits. We find that, in most cases, inserting a gate is the preferred option, whereas deleting a gate is almost never selected. Furthermore, the tree-search tends to add continuous gates (see details in \cref{sec:app_tree_search_add_gates}). Interestingly, the optimized angles of these newly added gates often place them close to an identity operation, suggesting that only small corrective adjustments are being applied.

\subsubsection{Ablation study}
\label{sec:ablation}

In this section, we demonstrate the impact of our proposed additions in \cref{sec:methods} to the overall model performance. Further details on the ablation studies are listed in \cref{sec:app_ablation}.

First, we vary the discrete noise schedule and keep the the schedule of the continuous mode fixed. In \cref{fig:model_ablation_scale}a, we observe our learned noise schedule lowers the infidelity significantly, compared to commonly used schedules: DDPM (Ref. \cite{ho2020denoising}) and Cosine (Ref. \cite{nichol2021improved}), showcasing the impact of an improper chosen schedule. Next, we switch on our proposed loss weighting which again improves the infidelity. Notably, we find our improvements consistently lower the infidelity across all tested model sizes.   
Additionally, we vary the continuous noise schedule while keeping the discrete schedule together with the loss weighting fixed. We observe that the infidelity does not change with statistical significance (see \cref{sec:app_ablation}), indicating that the main performance influence is attributed to the discrete mode.

\subsubsection{Scaling analysis}
\label{sec:scaling}

We now analyze the scaling potential of our method, differentiating between the scaling of the diffusion model and the unitary synthesis task. Here, we focus on the diffusion model and perform a scaling analysis for two model variables: the trainable parameter count (\cref{fig:model_ablation_scale}a), and the total number of training steps (\cref{fig:model_ablation_scale}b). In both cases, we see a steady improvement of the infidelity with increasing computation resources, following the scaling known for transformer-based DMs~\cite{peebles2023scalable}. We observe that the model benefits especially from more training steps, indicating that the correlations between gates and the unitary condition are very subtle and require many seen examples. Further, in \cref{fig:model_ablation_scale}b, we show that a model with 100M more parameters can be matched by a smaller one by training 200k update steps longer. 

\subsection{Hamiltonian evolution} 

\label{sec:hamiltonian}
We now evaluate the model's performance in a practical setting. We consider here the compilation of a unitary, $\mathcal{U}(\tau) = \exp(-i \tau H)$, that encodes the evolution of a system under a Hamiltonian $H$ for time $\tau$. This task, which involves decomposing the exponential operator into smaller operations, is commonly referred to as trotterization, and plays a central role in simulating quantum dynamics on quantum computers.

We consider here two paradigmatic: the Ising and Heisenberg XXZ models (see \cref{sec:app_ham}). In \cref{fig:hamiltonians}a, b, we show that the model generates accurate circuits across the phase space of both models and for different qubit counts. We observe that the infidelity slightly increases in regions of the phase space where the evolved state exhibits higher entanglement, causing the need to deeper circuits, following the trend observed in \cref{sec:benchmark}. In \cref{fig:hamiltonians}c, d, we display the infidelity as a function of $\tau$ for various points in the phase space. As before, larger values of $\tau$ typically result in more entangled states and, consequently, more complex circuits, which increases the infidelity. However, in some cases, the evolution induces an oscillatory behavior in the entanglement of the resulting state. For regions with lower entanglement the compilation is again more accurate, due to the need of shorter circuits. This effect is especially pronounced for the Ising model at $J\geq h$ and for the XXZ at $J, \Delta = 0.8$, where accuracy improves due to the lower circuit complexity.

\subsection{Identifying structures in generated circuits}

\begin{figure*}[ht!]
	\centering
	\includegraphics[width=\textwidth]{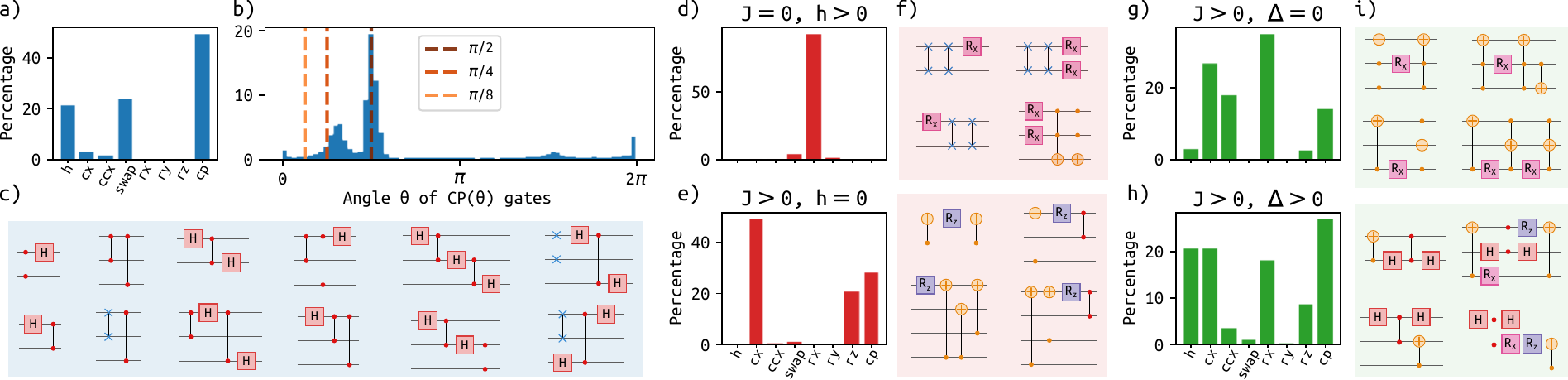}
	\caption{\textbf{Circuit structure analysis.}
        \textbf{a, b, c)} Gate distribution, angle distribution for the CP($\theta$) gates and most recurring gate sequences for circuits generated for the 4-qubit QFT.
        \textbf{d, e, f)} and \textbf{g, h, i)} Gate distribution for two points of the phase space (see titles) and most recurring gate sequences for the Ising (left) and XXZ (right) Hamiltonians, each for 4-qubits.
	}
	\label{fig:discovery}
\end{figure*}

\label{sec:discovery}
One of the advantages of the proposed model with respect to previous approaches is its sampling efficiency, enabling the fast generation of large numbers of candidate circuits for an input unitary. While not all of these circuits may have the desired infidelity, we find that the errors typically arise from the misplacement of a single or few gates, with the overall circuit structure being generally correct. In this section, we demonstrate how to exploit the full set of generated circuits for specific unitaries to uncover hidden structural patterns, gadgets, and distributions over continuous parameters. 

\paragraph{Quantum Fourier transform} The QFT is a quantum algorithm that exhibits an exponential advantage over its classical counterpart, the discrete Fourier transform~\cite{coppersmith2002approximate}. From a circuit perspective, the algorithm provides an efficient construction composed of three types of gates: Hadamard (H), Swap, and parameterized controlled-phase gates (CP($\theta$)).

When generating 2048 circuits from the QFT unitary, we obtain circuits with minimum infidelities of $1.4 \cdot 10^{-4}$, $3.8 \cdot 10^{-3}$, and $7.8 \cdot 10^{-2}$ for 3, 4 and 5 qubits, respectively. Moreover, we find that our model is able to generate circuits closely matching the textbook solution, unlike the compilers discussed in \cref{sec:comparison}, which typically rely on large optimized blocks of parameterized gates (see \cref{sec:app_comparison_compilers} and \cref{fig:app_additional_gadgets_qft_other_methods}). For the remainder of this analysis, however, we consider all generated circuits regardless of their infidelity. This allows us to analyze structural patterns without computing infidelities for each sample--a classically expensive operation--since we are interested in circuit structure rather than exact functional accuracy. Examining the histogram of gate types (\cref{fig:discovery}a), we find that the model predominantly predicts the expected gates. Furthermore, the distribution of predicted angles for the controlled-phase gate (\cref{fig:discovery}b) approximately matches the known target distribution (dashed lines). By performing GPE (see \cref{sec:gpe}), we are able to recover the building blocks of the standard QFT compilation protocol (see \cref{fig:discovery}c and \cref{fig:app_additional_gadgets_qft} for further examples). These findings suggest that, under the considered gate set, no alternative compilation strategies exist or may offer limited improvement over the canonical construction. 

\paragraph{Hamiltonian evolution} 
We now examine the circuits generated for the Hamiltonian evolutions introduced in \cref{sec:hamiltonian}. First, we focus on the Ising model.
At $J=0$ (\cref{fig:discovery}d), the Hamiltonian contains only the transverse-field term, consisting solely of single-qubit $X$ operators, that the model correctly compiles with single-qubit rotations ($R_x$) and Swap gates. Conversely, when $h=0$ (\cref{fig:discovery}e), the unitary reduces to a parameterized $Z\otimes Z$ interaction, which the model decomposes via CNOTs together with parameterized $R_z(\theta)$ and CP($\theta$). The gadgets extracted via GPE visually highlight the previous findings (\cref{fig:discovery}f). Notably, since the model is trained with a minimum of four gates, it tends to insert unnecessary Swap gates when the target unitary is close to the identity or corresponds to a single rotation, such as in the case of $h=0$. These redundant gates can, however, be removed with a straightforward optimization pass. For future works, we recommend adding 1 to 3 gate circuits into the training set, if one knows prior to the training such tiny circuits are required in the evaluation, mitigating this bias.

Second, we analyze the circuits generated for the XXZ model. Here, we fix the transverse field at $h=0.2$ and vary the coupling $J$ and $\Delta$ (see \cref{sec:app_ham}). For $\Delta = 0$, the Hamiltonian contains an $X\otimes X + Y\otimes Y$ interaction, known in quantum computing as an iSWAP gate. Because iSWAP is not in our gate set, the model instead decomposes this interaction using CNOTs, Toffoli and CP($\theta$) gates (\cref{fig:discovery}g), which interestingly differs from the usual decomposition with $R_z(\pi /2)$, CNOT and H gates~\cite{schuch2003natural}. On the other hand, when both $J\neq0$ and $\Delta\neq0$ , a $Z\otimes Z$ term enters the dynamics, and we observe a marked increase in the usage of H and CP($\theta$) gates, which surprisingly varies from this term’s decomposition in the Ising case (\cref{fig:discovery}h). Again, by inspecting the GPE (\cref{fig:discovery}i), we observe clearly the striking differences between the properties of the same Hamiltonian in two different points of its phase space.

\section{Discussion}

In this work we introduced a multimodal diffusion model that jointly synthesizes the gate sequence and continuous parameters of a quantum circuit. To do so, we leverage two independent diffusion processes, one for the discrete mode and one for the continuous mode. This separation lets each mode be sampled and potentially improved in isolation. For example, we envision that masked diffusion and autoregressive decoding could further improve the discrete mode prediction. The use of diffusion models for quantum circuit synthesis, first introduced in Ref.~\cite{furrutter2024quantum}, opens the door to efficient unitary synthesis. The present work builds on that foundation by enhancing the model architecture and extending it to support continuous gates. Compared to state-of-the-art compilers, our method generates shorter circuits with significantly lower runtimes, though this comes at the cost of reduced compilation accuracy. However, such a disadvantage only holds in a regime of synthesizing larger-scale circuits and in the context of fault-tolerant quantum computing. When considering typical noise regimes of NISQ devices, the ability of genQC2 to compile shorter circuits leads to much larger compilation accuracies (see \cref{fig:noisy_simulation}). 

Most importantly, the method aims at performing \emph{exact} compilation, as opposed to approximate compilation done by existing methods. Rather than stacking blocks of parametrized gates and optimizing these, the model creates more natural structures, highlighted for instance by the finding of the textbook compilation of the quantum Fourier transform (see \cref{sec:discovery}). Such ability, combined with other methods such as gadget mining~\cite{trenkwalder2023automated, kundu2024easy,olle2025scaling} or the proposed gate-pair encodings (\cref{sec:gpe}) can be leveraged to discovery new compilation heuristics and generally to better quantum circuit synthesis.

From an application perspective, although we focused on unitary synthesis, the same pipeline can be adapted to tasks such as state preparation, eigensolvers, error-correction decoding, or circuit design for photonic and measurement-based platforms. In particular, a promising direction is to fine-tune the current model for any of these tasks. By applying contrastive learning with task-specific loss functions, such as those used in VQE, one can adapt the model to generate circuits either for new prompts or tailored to the general unitary distribution of the problem at hand. Notably, as the condition is arbitrary, the method can also be applied to non-unitary matrices, enabling the compilation of more advanced quantum algorithms. 

Two main limitations remain: accuracy and scalability. On the former, we have shown that the model’s fidelity still trails full “search-plus-gradient” pipelines~\cite{smith2023LEAP,  madden2022best,grimsley2019adaptive}, and its performance degrades with increasing gate counts (\cref{sec:benchmark}). We expect that better DM training protocols and sampling methods will close much of this gap. Recent works on discrete DMs, and specifically on Masked DMs~\cite{sahoo2024simple, shi2024simplified}, are promising upgrades for the architecture, which, due to its multimodal architecture, could be adapted swiftly. Moreover, existing pipelines can leverage the sampling efficiency of the model, proposing already accurate candidate circuits from which further optimization can be performed (see \cref{sec:gen_ansatz}). On the other hand, the quantum nature of the problem makes scalability an important bottleneck. For instance, the input unitary matrix grows exponentially with qubit count. Future work should explore better conditioning, as for instance directly inputting the Hamiltonians used in~\cref{sec:hamiltonian} in text form or similar symbolic descriptions. Moreover, compared with DMs that handle images consisting of thousands of pixels, our circuits are much smaller in token count, i.e. order of (max. \#qubits x max. \#gates) = 5x32, so scaling the architecture is plausible, given one can sample a sufficiently large dataset. The real challenge is conceptual: deeper circuits encode harder quantum tasks (especially as the circuit space explodes exponentially with the gate count) and will likely require smarter representations, e.g., compressing the dataset with the gate-pair encodings discussed in~\cref{sec:discovery} to lower the number of tokens per circuit  or similar approaches~\cite{trenkwalder2023automated, kundu2024easy,olle2025scaling}.


\section*{Acknowledgments}

This research was funded in part by the Austrian Science Fund (FWF) [SFB BeyondC F7102, 10.55776/F71]. For open access purposes, the author has applied a CC BY public copyright license to any author accepted manuscript version arising from this submission. This work was also supported by the European Union (ERC Advanced Grant, QuantAI, No. 101055129). The views and opinions expressed in this article are however those of the author(s) only and do not necessarily reflect those of the European Union or the European Research Council - neither the European Union nor the granting authority can be held responsible for them. This research used resources of the National Energy Research Scientific Computing Center (NERSC), a Department of Energy User Facility using NERSC award ERCAP0032002.

\section*{Code and data availability}

All the resources necessary to reproduce the results in this paper are accessible in Ref.~\cite{github}. The code is given in the form of a Python library, \verb|genQC|, which allows the user to train new models or generate circuits from pre-trained models. The library also contains multiple tutorials that will guide the user through the various applications of the proposed method. 
The training dataset is not shared due to memory constraints, but can be generated with the released code. All necessary details are provided in \cref{sec:methods}, \cref{sec:experiments} and the Supplementary Material.


\bibliography{biblio.bib} 

\onecolumngrid 

\include{appendix.tex}

\end{document}

%% file: appendix.tex
\clearpage
\appendix

\setcounter{page}{1}

\begin{center}
	\LARGE\bfseries{\MyPaperTitle}\par
	\vspace{0.2cm}
	\Large{Supplementary Material} \par
\end{center}
\begin{center}\makebox[\columnwidth]{\rule{0.8\columnwidth}{0.1pt}}\end{center}

\begin{figure}[h!]
	\centering
	\includegraphics[width=0.95\columnwidth]{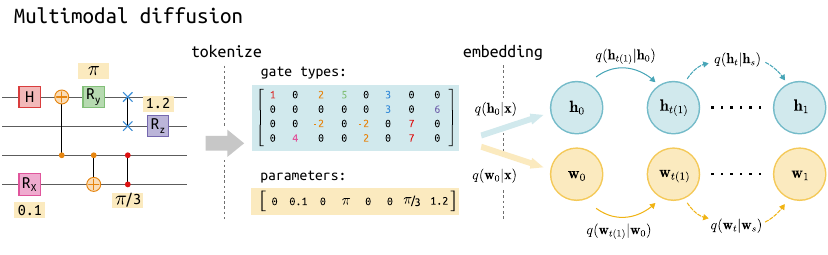}
	\caption{\textbf{Multimodal diffusion.}
		Overview of the tokenization, embedding and forward diffusion process of our pipeline. See \cref{sec:app_multimode_details} and \cref{sec:app_emb} for details.
	}\label{fig:app_DM_overview}
\end{figure}

\section{Multimodal diffusion details}\label{sec:app_multimode_details}
In this section, we detail the design of our pipeline for the generation of quantum circuits, which consist of discrete (categorical) and continuous data. Generally, we use the notation of discrete time diffusion for times $t(i)=i/T$ and $s(i)=(i-1)/T$ such that $0\leq s<t \leq1$, where $T$ is the number of timesteps and $i$ a time index running from 1 to $T$. Further, we assume that $\vb{x}$ denotes a quantum circuit with its embedding split into a token embedding $\vb{h}_0$, encoding the gate type, and a continuous embedding $\vb{w}_0$, describing the angle of the parameterized gates (see \cref{fig:DM_overview}a and \cref{fig:app_DM_overview}).

As described in \cref{sec:multidiff} and \cref{eq:forward_multimodial}, we model the forward diffusion as two independent Gaussian diffusion processes
\begin{equation*}
	q(\vb{h}_{0:1}, \vb{w}_{0:1} | \vb{x}) = q(\vb{h}_{0:1} | \vb{x})\; q(\vb{w}_{0:1} | \vb{x}),
\end{equation*}
The joint forward  distributions are the usual discrete-time Markov chains~\cite{ho2020denoising, kingma2021variational}, defined as
\begin{align}
		q(\vb{h}_{0:1} | \vb{x}) &= q(\vb{h}_{1})\;q(\vb{h}_{0} | \vb{x})\prod_{i=1}^{T_h} q(\vb{h}_{t(i)}|\vb{h}_{s(i)}) \quad\text{and} \\
		q(\vb{w}_{0:1} | \vb{x}) &= q(\vb{w}_{1})\;q(\vb{w}_{0} | \vb{x})\prod_{i=1}^{T_w} q(\vb{w}_{t(i)}|\vb{w}_{s(i)}),
\end{align}
where $q(\vb{h}_{1})=\mathcal{N}(\vb{0}, \vb{I})$ and $q(\vb{w}_{1})=\mathcal{N}(\vb{0}, \vb{I})$. Further, we set $T_h=T_w=T$ and for each mode consider fixed deterministic encoders $q(\vb{h}_0|\vb{x})$ and $q(\vb{w}_0|\vb{x})$. Following from the definition \cref{eq:gaussian_dm}, the Markov transitions are specified by 
\begin{align}
	q(\vb{h}_{t(i)}|\vb{h}_{s(i)}) &= \mathcal{N}(\vb{h}_{t(i)}; \sqrt{\alpha_{t(i)}^h}\vb{h}_{s(i)}, \beta_{t(i)}^h\vb{I}) \label{eq:app_markov_transitions1} \quad\text{and} \\
	q(\vb{h}_{t(i)}|\vb{h}_0) &= \mathcal{N}(\vb{h}_{t(i)}; \sqrt{\bar{\alpha}_{t(i)}^h}\vb{h}_0, \bar{\beta}_{t(i)}^h\vb{I}). \label{eq:app_markov_transitions2}
\end{align}
The same approach is then followed for $q(\vb{w}_{t(i)}|\vb{w}_{s(i)})$ and $q(\vb{w}_{t(i)}|\vb{w}_0)$ with their respective parameters $\alpha_{t(i)}^w$, $\beta_{t(i)}^w$ and $\bar{\alpha}_{t(i)}^w$, $\bar{\beta}_{t(i)}^w$. Additionally, \cref{eq:app_markov_transitions1} and \cref{eq:app_markov_transitions2} are related by $\bar{\alpha}_{t(i)}=\prod_{k=0}^{i}\alpha_{t(k)}$ and $\bar{\beta}_{t(i)}=\prod_{k=0}^{i}\beta_{t(k)}=\prod_{k=0}^{i}\br{1-\alpha_{t(k)}}$.

\paragraph{Generative model}
Given a condition $\vb{c}$, we set our conditional generative model $p_\theta(\vb{x}, \vb{h}_{0:1}, \vb{w}_{0:1}|\vb{c})$ to a general formulation:
\begin{equation}\label{eq:joint_model}
	p_\theta(\vb{x}, \vb{h}_{0:1}, \vb{w}_{0:1}|\vb{c}) = p(\vb{x}| \vb{h}_{0}, \vb{w}_{0})\; p_\theta(\vb{h}_{0:1}, \vb{w}_{0:1}|\vb{c}),
\end{equation}
where $p(\vb{x}| \vb{h}_{0}, \vb{w}_{0})$ is the embedding decoder transforming the latent representation back to the circuit picture, and $p_\theta(\vb{h}_{0:1},\vb{w}_{0:1}|\vb{c})$ depends on our sampling approach (see \cref{sec:app_sample_modes}). Following Ref.~\cite{bao2023one}, we define a reverse transition model that accounts for: (i) all marginal distributions $p_\theta(\vb{h}_{0:1}|\vb{c})$ and $p_\theta(\vb{w}_{0:1}|\vb{c})$; (ii) all conditional distributions $p_\theta(\vb{h}_{0:1}|\vb{w}_{0},\vb{c})$ and $p_\theta(\vb{w}_{0:1}|\vb{h}_{0}, \vb{c})$; (iii) and the joint $p_\theta(\vb{h}_{0:1},\vb{w}_{0:1}|\vb{c})$. Specifically, 
\begin{equation}\label{eq:unsplit_gen_model}
	p_\theta(\vb{h}_{s}, \vb{w}_ {\tilde{s}}|\vb{h}_{t}, \vb{w}_{\tilde{t}}, \vb{c}) = q(\vb{h}_{s}, \vb{w}_{\tilde{s}}|\vb{h}_{t}, \vb{w}_{\tilde{t}}, \vb{x}=\hat{\vb{x}}_\theta(\vb{h}_{t}, \vb{w}_{\tilde{t}}, t, \tilde{t}, \vb{c})),
\end{equation}
where $0\leq s<t \leq1$ and $0\leq \tilde{s}<\tilde{t} \leq1$ are independent diffusion times, and $q(\vb{h}_{s}, \vb{w}_{\tilde{s}}|\vb{h}_{t}, \vb{w}_{\tilde{t}}, \vb{x})$ is the true top-down posterior given a data sample $\vb{x}$ (see \cref{sec:app_top_down_posterior} for the exact derivation). Here, we note that the diffusion model $\hat{\vb{x}}_\theta$ predicts the denoised circuit $\vb{x}$ implicitly by predicting both modes (see \cref{fig:DM_overview}b). This is possible as the top-down posterior splits up for a given $\vb{x}$ due to the independent forward diffusion, allowing us to rewrite 
\begin{equation}\label{eq:split_gen_model}
	p_\theta(\vb{h}_{s}, \vb{w}_{\tilde{s}}|\vb{h}_{t}, \vb{w}_{\tilde{t}}, \vb{c}) = p_\theta(\vb{h}_{s}|\vb{h}_{t}, \vb{w}_{\tilde{t}}, \vb{c}) \; p_\theta(\vb{w}_ {\tilde{s}}|\vb{w}_{\tilde{t}}, \vb{h}_{t}, \vb{c}),
\end{equation}
where the model captures the correlations between $\vb{h}_{s}$ and $\vb{w}_{\tilde{s}}$ by jointly predicting $[\vb{h}_{0}(\vb{x}), \vb{w}_{0}(\vb{x})] = \hat{\vb{x}}_{\theta}(\vb{h}_{t}, \vb{w}_{\tilde{t}}, t, \tilde{t}, \vb{c})$. Following the independent forward diffusion in \cref{eq:forward_multimodial}, the joint distribution at the final timestep $t=1$ is the product of two independent standard normal distributions, $p(\vb{h}_{1}, \vb{w}_{1}) = p(\vb{h}_{1})\;  p(\vb{w}_{1})$. In addition, we sample the marginal distributions in (i) by
conditioning on the fully noisy version of the other mode. For instance, $p_\theta(\vb{h}_{0:1}|\vb{c})$ is sampled via $p_\theta(\vb{h}_{s}|\vb{h}_{t}, \vb{w}_{1}, \vb{c})$ and vice versa. On the other hand, following the classifier-free guidance (CFG) approach~\cite{ho2022classifier}, the condition $\vb{c}$ is randomly replaced with an empty condition $\phi$ during training with a fixed probability, enabling condition-free sampling.

\subsection{Generative model sampling modes}\label{sec:app_sample_modes}
Here, we show how to define the generative model $p_\theta(\vb{h}_{0:1},\vb{w}_{0:1}|\vb{c})$ of \cref{eq:joint_model}, using the appropriate reverse transitions stemming from \cref{eq:unsplit_gen_model} and \cref{eq:split_gen_model}, as explained in the main text. For the sake of clarity, we drop the condition on $\vb{c}$.

The multimodal model can be sampled in multiple valid ways:
\begin{enumerate}[(i)]
	\item{\textbf{Joint modes}}: Assuming that we want to sample jointly the gate types and the continuous parameters, we write
	\begin{align}
		p_\theta(\vb{h}_{0:1},\vb{w}_{0:1})
		&= p(\vb{h}_{1}, \vb{w}_{1}) \prod_{i=1}^{T} p_\theta(\vb{h}_{s(i)}, \vb{w}_ {\tilde{s}(i)}|\vb{h}_{t(i)}, \vb{w}_{\tilde{t}(i)}) \nonumber\\
		&= p(\vb{h}_{1}) \; p(\vb{w}_{1})\prod_{i=1}^{T} p_\theta(\vb{h}_{s(i)}|\vb{h}_{t(i)}, \vb{w}_{\tilde{t}(i)}) \; p_\theta(\vb{w}_{\tilde{s}(i)}|\vb{w}_{\tilde{t}(i)},
		 \vb{h}_{t(i)}). \label{eq:app_joint_modes}
	\end{align}
	
	\item{\textbf{Sequential modes}}: Equivalently to \cref{eq:app_joint_modes}, we can sample first the discrete tokens $\vb{h}_0$ independently, and sequentially generate the corresponding parameters $\vb{w}_0$ by
	\begin{align}
		p_\theta(\vb{h}_{0:1},\vb{w}_{0:1})
		&= p_\theta(\vb{h}_{0:1})\; p_\theta(\vb{w}_{0:1}|\vb{h}_{0}) \nonumber\\
		&= p(\vb{h}_{1}) \prod_{i=1}^{T} p_\theta(\vb{h}_{s(i)}|\vb{h}_{t(i)}) \;\cdot\;
		p(\vb{w}_{1}) \prod_{j=1}^{T} p_\theta(\vb{w}_{\tilde{s}(j)}|\vb{w}_{\tilde{t}(j)}, \vb{h}_{0}). \label{eq:app_sequential_modes}
	\end{align}
	
	\item{\textbf{Single mode}}: Building upon \cref{eq:app_sequential_modes}, we can also assume that we are given a gate Ansatz, which means a given, fixed  $\vb{h}_0$. If we want to generate the corresponding parameters, we can then just use the second part of previous equation, namely,
	\begin{equation}
		p(\vb{w}_{0:1}|\vb{h}_0) = p(\vb{w}_{1}) \prod_{j=1}^{T} p_\theta(\vb{w}_{\tilde{s}(j)}|\vb{w}_{\tilde{t}(j)}, \vb{h}_{0}). \label{eq:app_single_mode}
	\end{equation}
	
	\item{\textbf{Independent modes}}: Finally, the use of classifier-free guidance (CFG)~\cite{ho2022classifier} requires the definition of unconditional marginals, i.e. $p(\vb{h}_0)$ and $p(\vb{w}_0)$, which here we can sampled using
	\begin{align}
		p_\theta(\vb{h}_{0:1}) &= p(\vb{h}_{1}) \prod_{i=1}^{T} p_\theta(\vb{h}_{s(i)}|\vb{h}_{t(i)}) 
		\quad\text{and} \\
		p_\theta(\vb{w}_{0:1}) &= p(\vb{w}_{1}) \prod_{i=1}^{T} p_\theta(\vb{w}_{s(i)}|\vb{w}_{t(i)}). 
	\end{align}	
\end{enumerate}
As (ii) requires double the number of model evaluations, we only use (i) for joint sampling in \cref{sec:experiments}.

\subsection{Multimodal top-down posterior}\label{sec:app_top_down_posterior}
We show here that the top-down posterior from \cref{eq:unsplit_gen_model} splits to \cref{eq:split_gen_model}. Considering $0\leq s<t \leq1$ and $0\leq \tilde{s}<\tilde{t} \leq1$, we analyze the multimodal posterior by splitting the joint distribution
\begin{equation}\label{eq:app_raw_split}
	q(\vb{h}_{s}, \vb{w}_{\tilde{s}}|\vb{h}_{t}, \vb{w}_{\tilde{t}}, \vb{x})
	= 
	q(\vb{h}_{s} | \vb{h}_{t}, \vb{w}_{\tilde{t}}, \vb{x}) \;
	q(\vb{w}_{\tilde{s}} | \vb{h}_{s}, \vb{h}_{t}, \vb{w}_{\tilde{t}}, \vb{x}).
\end{equation}
We can simplify the second term of the previous equation using Bayes' law and the Markov property of the diffusion process, finding
\begin{align}
	q(\textcolor{magenta}{\vb{w}_{\tilde{s}}} | \vb{h}_{s}, \textcolor{cyan}{\vb{h}_{t}}, \vb{w}_{\tilde{t}}, \vb{x}) 
	&= \frac{q(\textcolor{cyan}{\vb{h}_{t}} | \vb{h}_{s}, \textcolor{magenta}{\vb{w}_{\tilde{s}}}, \vb{w}_{\tilde{t}}, \vb{x})}{q(\textcolor{cyan}{\vb{h}_{t}} | \vb{h}_{s}, \vb{w}_{\tilde{t}}, \vb{x})}
	\; q(\textcolor{magenta}{\vb{w}_{\tilde{s}}} | \vb{h}_{s}, \vb{w}_{\tilde{t}}, \vb{x}) \nonumber     \\
	&= \frac{q(\textcolor{cyan}{\vb{h}_{t}} | \vb{h}_{s})}{q(\textcolor{cyan}{\vb{h}_{t}} | \vb{h}_{s})} q(\textcolor{magenta}{\vb{w}_{\tilde{s}}} | \vb{h}_{s}, \vb{w}_{\tilde{t}}, \vb{x}) \nonumber     \\
	&= q(\textcolor{magenta}{\vb{w}_{\tilde{s}}} | \vb{h}_{s}, \vb{w}_{\tilde{t}}, \vb{x}). \label{eq:app_app_raw_split1}
\end{align}
In the previous equation, the second equality was done by means of the forward process \cref{eq:forward_multimodial}, that allows us to derive
\begin{align}\label{eq:app_app_raw_split11}
	q(\textcolor{cyan}{\vb{h}_{t}} | \vb{h}_{s}, \textcolor{magenta}{\vb{w}_{\tilde{s}}}, \vb{w}_{\tilde{t}}, \vb{x}) 
	&= \frac{q(\textcolor{cyan}{\vb{h}_{t}}, \vb{h}_{s}, \textcolor{magenta}{\vb{w}_{\tilde{s}}}, \vb{w}_{\tilde{t}}, \vb{x})}{q(\vb{h}_{s}, \textcolor{magenta}{\vb{w}_{\tilde{s}}}, \vb{w}_{\tilde{t}}, \vb{x}) }
	\nonumber     \\
	&= \frac{q(\textcolor{cyan}{\vb{h}_{t}}, \vb{h}_{s} | \vb{x}) \; q( \textcolor{magenta}{\vb{w}_{\tilde{s}}}, \vb{w}_{\tilde{t}} | \vb{x})}{q(\vb{h}_{s}| \vb{x}) \; q( \textcolor{magenta}{\vb{w}_{\tilde{s}}}, \vb{w}_{\tilde{t}}| \vb{x}) }
	\nonumber     \\
	&= \frac{q(\textcolor{cyan}{\vb{h}_{t}}| \vb{h}_{s} , \vb{x}) \; q(\vb{h}_{s} | \vb{x})}{q(\vb{h}_{s}| \vb{x}) }
	\nonumber     \\
	&= q(\textcolor{cyan}{\vb{h}_{t}}| \vb{h}_{s}),
\end{align}
A similar derivation can be done for the term $q(\textcolor{cyan}{\vb{h}_{t}} | \vb{h}_{s}, \vb{w}_{\tilde{t}}, \vb{x})$  in the denominator in \cref{eq:app_app_raw_split1}. We are using similar variants of \cref{eq:app_app_raw_split11} below, i.e. for $q(\textcolor{teal}{\vb{h}_{s}}| \textcolor{magenta}{\vb{w}_{\tilde{s}}}, \vb{w}_{\tilde{t}}, \vb{x})$ and $q(\textcolor{teal}{\vb{h}_{s}}| \vb{w}_{\tilde{t}}, \vb{x})$ in \cref{eq:app_app_raw_split2}, and for $q(\textcolor{Bittersweet}{\vb{w}_{\tilde{t}}} | \vb{h}_{t}, \textcolor{teal}{\vb{h}_{s}}, \vb{x})$ and $q(\textcolor{Bittersweet}{\vb{w}_{\tilde{t}}} | \vb{h}_{t}, \vb{x})$ in \cref{eq:app_app_raw_split3} . Further, we simplify \cref{eq:app_app_raw_split1} by
\begin{align}
	q(\textcolor{magenta}{\vb{w}_{\tilde{s}}} | \textcolor{teal}{\vb{h}_{s}}, \vb{w}_{\tilde{t}}, \vb{x})
	&= \frac{q(\textcolor{teal}{\vb{h}_{s}}| \textcolor{magenta}{\vb{w}_{\tilde{s}}}, \vb{w}_{\tilde{t}}, \vb{x})}{q(\textcolor{teal}{\vb{h}_{s}}| \vb{w}_{\tilde{t}}, \vb{x})}
	q(\textcolor{magenta}{\vb{w}_{\tilde{s}}} | \vb{w}_{\tilde{t}}, \vb{x}) \nonumber     \\
	&= \frac{q(\textcolor{teal}{\vb{h}_{s}}| \vb{x})}{q(\textcolor{teal}{\vb{h}_{s}}| \vb{x})}
	q(\textcolor{magenta}{\vb{w}_{\tilde{s}}} | \vb{w}_{\tilde{t}}, \vb{x}) \nonumber     \\
	&= q(\textcolor{magenta}{\vb{w}_{\tilde{s}}} | \vb{w}_{\tilde{t}}, \vb{x}), \label{eq:app_app_raw_split2}
\end{align}
which is the usual top-down posterior for $\vb{w}_{\tilde{s}}$ for standard Gaussian DMs, as used for instance in DDPM~\cite{ho2020denoising}. In the same direction as done above, for the top-down posterior of $\vb{h}_{s}$  (the first part of \cref{eq:app_raw_split}) we derive
\begin{align}
	q(\textcolor{teal}{\vb{h}_{s}} | \vb{h}_{t}, \textcolor{Bittersweet}{\vb{w}_{\tilde{t}}}, \vb{x})
	&= \frac{q(\textcolor{Bittersweet}{\vb{w}_{\tilde{t}}} | \vb{h}_{t}, \textcolor{teal}{\vb{h}_{s}}, \vb{x})}{q(\textcolor{Bittersweet}{\vb{w}_{\tilde{t}}} | \vb{h}_{t}, \vb{x})} q(\textcolor{teal}{\vb{h}_{s}} | \vb{h}_{t}, \vb{x}) \nonumber     \\
	&= \frac{q(\textcolor{Bittersweet}{\vb{w}_{\tilde{t}}} |  \vb{x})}{q(\textcolor{Bittersweet}{\vb{w}_{\tilde{t}}} | \vb{x})} q(\textcolor{teal}{\vb{h}_{s}} | \vb{h}_{t}, \vb{x}) \nonumber     \\
	&= q(\textcolor{teal}{\vb{h}_{s}} | \vb{h}_{t}, \vb{x}). \label{eq:app_app_raw_split3}
\end{align}
Now, using \cref{eq:app_app_raw_split1}, \cref{eq:app_app_raw_split2} and \cref{eq:app_app_raw_split3}, we rewrite \cref{eq:app_raw_split} to 
\begin{align}	
	q(\vb{h}_{s}, \vb{w}_{\tilde{s}}|\vb{h}_{t}, \vb{w}_{\tilde{t}}, \vb{x})
	&= q(\vb{h}_{s} | \vb{h}_{t}, \vb{w}_{\tilde{t}}, \vb{x}) \;
	q(\vb{w}_{\tilde{s}} | \vb{h}_{s}, \vb{h}_{t}, \vb{w}_{\tilde{t}}, \vb{x}) \nonumber\\
	&= q(\vb{h}_{s} | \vb{h}_{t}, \vb{w}_{\tilde{t}}, \vb{x}) \;
	q(\vb{w}_{\tilde{s}} | \vb{h}_{s}, \vb{w}_{\tilde{t}}, \vb{x})\nonumber\\
	&= q(\vb{h}_{s} | \vb{h}_{t}, \vb{x}) \;
	q(\vb{w}_{\tilde{s}} | \vb{w}_{\tilde{t}}, \vb{x}). \label{eq:app_split_top_down}		
\end{align}
The \cref{eq:app_split_top_down} proves the splitting of the reverse transitions of \cref{eq:split_gen_model}. Specifically, we set for the model
\begin{align}\label{eq:app_split_gen_model}
	p_\theta(\vb{h}_{s}, \vb{w}_{\tilde{s}}|\vb{h}_{t}, \vb{w}_{\tilde{t}}, \vb{c}) 
	&= p_\theta(\vb{h}_{s}|\vb{h}_{t}, \vb{w}_{\tilde{t}}, \vb{c}) \; p_\theta(\vb{w}_ {\tilde{s}}|\vb{w}_{\tilde{t}}, \vb{h}_{t}, \vb{c}) \nonumber\\
	&= q(\vb{h}_{s}|\vb{h}_{t}, \hat{\vb{x}}_\theta(\vb{h}_{t}, \vb{w}_{\tilde{t}}, t, \tilde{t}, \vb{c})) \;
	q(\vb{w}_ {\tilde{s}}|\vb{w}_{\tilde{t}}, \hat{\vb{x}}_\theta(\vb{h}_{t}, \vb{w}_{\tilde{t}}, t, \tilde{t}, \vb{c})).
\end{align}

\subsection{Multimodal ELBO} \label{sec:app_multi_elbo}
In order to train the diffusion model defined in the main text and above, we optimize the log-likelihood of $\vb{x}$ for the multimodal diffusion model with the evidence lower bound (ELBO), which can be derived as
\begin{align}	
	\log p(\vb{x}) &= \log \int p_\theta(\vb{x}, \vb{h}_{0:1}, \vb{w}_{0:1}) \dd{\vb{h}_{0:1}}\dd{\vb{w}_{0:1}} \nonumber\\
	&= \log \int \frac{p_\theta(\vb{x}, \vb{h}_{0:1}, \vb{w}_{0:1})\; q(\vb{h}_{0:1}, \vb{w}_{0:1} | \vb{x})}{q(\vb{h}_{0:1}, \vb{w}_{0:1} | \vb{x})} \dd{\vb{h}_{0:1}}\dd{\vb{w}_{0:1}} \nonumber\\
	&= \log\Expec{q(\vb{h}_{0:1}, \vb{w}_{0:1} | \vb{x})}{\frac{p_\theta(\vb{x}, \vb{h}_{0:1}, \vb{w}_{0:1})}{q(\vb{h}_{0:1}, \vb{w}_{0:1} | \vb{x})}} \nonumber\\
	&\geq \Expec{q(\vb{h}_{0:1}, \vb{w}_{0:1} | \vb{x})}{\log\frac{p_\theta(\vb{x}, \vb{h}_{0:1}, \vb{w}_{0:1})}{q(\vb{h}_{0:1}, \vb{w}_{0:1} | \vb{x})}}.
\end{align}
The loss $\mathcal{L}(\vb{x})$ used for training is then defined as
\begin{equation}
	\mathcal{L}(\vb{x}) 
	\coloneq -\Expec{q(\vb{h}_{0:1}, \vb{w}_{0:1} | \vb{x})}{\log\frac{p(\vb{x}| \vb{h}_{0}, \vb{w}_{0})\; p_\theta(\vb{h}_{0:1}, \vb{w}_{0:1})}{q(\vb{h}_{0:1} | \vb{x})\; q(\vb{w}_{0:1} | \vb{x})}},
\end{equation}
which reduces to the loss of the main paper (see \cref{eq:loss}) when accounting for: 1) all sampling modes from \cref{sec:app_sample_modes}; 2) the use of the multimodal top-down posterior of \cref{sec:app_top_down_posterior}.

\subsection{Velocity loss weight}\label{sec:app_v_weight}

As described in \cref{sec:multidiff}, we use the velocity target~\cite{salimansprogressive} 
\begin{equation}\label{eq:app_velocity_def}
	\vb{v}_t \coloneq \sqrt{\bar{\alpha}_t}\,\vb*{\epsilon}-\sqrt{1-\bar{\alpha}_t}\,\vb{x},
\end{equation}
ensuring that the model outputs have unit variance. In contrast, the rectified flow-based velocity~\cite{liu2023flow}, defined as $\vb{v}_{\text{flow}}\coloneq\vb*{\epsilon}-\vb{x}$, does not guarantee this property. For a Gaussian diffusion process (see \cref{eq:gaussian_dm}), we sample a latent $\vb{z}_t$ at time $t$ using
\begin{equation}\label{eq:app_forward_sample}
	\vb{z}_t = \sqrt{\bar{\alpha}_t}\,\vb{x} + \sqrt{1-\bar{\alpha}_t}\,\vb*{\epsilon},
\end{equation}
where $\vb*{\epsilon}\sim\mathcal{N}(\vb{0}, \vb{I})$. Next, we combine \cref{eq:app_velocity_def} and \cref{eq:app_forward_sample} to 
\begin{equation}
	\vb{x} = \sqrt{\bar{\alpha}_t}\,\vb{z}_t - \sqrt{1-\bar{\alpha}_t}\,\vb{v}_t \label{eq:app_x_from_v}.
\end{equation}

Next, we derive the conversion factor $\omega_x(t)=(\mathrm{SNR}(t)+1)\,\omega_v(t)$, mentioned in \cref{sec:multidiff}, that arises when one translates the loss from a denoising prediction $\hat{\vb{x}}_{\theta}$ to a velocity predictor $\hat{\vb{v}}_{\theta}$. For this, we insert \cref{eq:app_x_from_v} into the MSE 
\begin{align}
	\omega_x(t)\norm{\vb{x} - \hat{\vb{x}}_{\theta}}_2^2
	&= \omega_x(t)\norm{\sqrt{1-\bar{\alpha}_t} \br{\vb{v}_t - \hat{\vb{v}}_{\theta}}}_2^2
	\nonumber\\
	&\coloneq \omega_v(t)\norm{ \br{\vb{v}_t - \hat{\vb{v}}_{\theta}}}_2^2,
\end{align}
where one can show for a variance-preserving diffusion process the relation $(\mathrm{SNR}(t)+1)^{-1}=(1-\bar{\alpha}_{t})$. Hence, we get the final relation
\begin{equation}
	\omega_v(t)=(1-\bar{\alpha}_{t})\,\omega_x(t),
\end{equation}
which we use for the loss weights in \cref{eq:loss}, as introduced in the main text.

\subsection{Signal-to-noise ratio for continuous parameters}\label{sec:app_cont_snr}

As mentioned above, the forward process of the continuous embeddings $\vb{w}_t$ is given as
\begin{equation}\label{eq:app_cont_snr_forward}
	q(\vb{w}_{t}|\vb{w}_0) = \mathcal{N}(\vb{w}_{t}; \sqrt{\bar{\alpha}_{t}^w}\vb{w}_0, \bar{\beta}_{t}^w\vb{I}),
\end{equation}
where $\bar{\beta}_{t}^w=1-\bar{\alpha}_{t}^w$. Moreover, we consider an embedding of the parameter $\lambda\in[-1,1]$ via (see \cref{sec:embedding} and \cref{sec:app_cont_emb})
\begin{equation}\label{eq:app_cont_snr_forward2}
	\vb{w}_0 = \cos(\lambda\pi) \vb{v}_1 + \sin(\lambda\pi) \vb{v}_2.
\end{equation}

As explained in \cref{sec:multidiff}, we use a sigmoid log-SNR weighting from Ref.~\cite{hoogeboom2024simpler}
\begin{equation}\label{eq:app_sigmoid_weight}
	\omega_w(t)=\mathrm{sigmoid}\bre{\log(\mathrm{SNR}(t))}.
\end{equation}
Notably, due to the encoding of the parameter $\lambda$ (see \cref{eq:app_cont_snr_forward2} and \cref{sec:app_cont_emb}), we have to alter the normal SNR in \cref{eq:app_sigmoid_weight}, initially given by $\mathrm{SNR}_w(t)\coloneq\bar{\alpha}_t^w / (1-\bar{\alpha}_t^w)$. This redefinition is required because the signal is now encoded into two correlated dimensions, i.e. $\vb{v}_1$ and $\vb{v}_2$, and is therefore more robust to the noise of \cref{eq:app_cont_snr_forward}. The new $\mathrm{SNR}_\lambda$ for the continuous parameters $\lambda$, with its noisy reconstruction $\hat{\lambda}_t$ (see \cref{sec:app_cont_emb}), is defined as
\begin{equation}\label{eq:app_cont_snr}
	\mathrm{SNR}_\lambda(t) \coloneq \frac{\Expec{}{\lambda^2}}{\varOp\hat{\lambda}_t} = \frac{\text{const.}}{\varOp\hat{\lambda}_t}.
\end{equation}
Here, we assume that the samples of $\lambda$ from the prior $p(\lambda)$ are bounded and normalized. Since $\lambda$ is periodic in $[ -1, 1 ]$ the absolute value is not relevant and we treat $\Expec{}{\lambda^2}$ as a constant, as it does not depend on the diffusion time $t$. Indeed, we can view it as a constant shift for the sigmoid weighting in \cref{eq:app_sigmoid_weight}. 

Given the previous definition, we can get an upper bound of \cref{eq:app_cont_snr} by using the \textit{Cramér–Rao bound}~\cite{cramer1946mathematical} for an estimator of the parameter $\lambda$,
\begin{equation}
	\frac{1}{\varOp\hat{\lambda}_t} \leq \mathcal{I}(\lambda),
\end{equation}
where $\mathcal{I}(\cdot)$ is the Fisher information. We do so by first, calculating the score function $s(\lambda, \vb{w}_t)$ of \cref{eq:app_cont_snr_forward}
\begin{align}
	s(\lambda, \vb{w}_t) &= \pdv{\lambda} \log(p(\vb{w}_t|\lambda)) \\
	&= - \pdv{\lambda} \frac{1}{2\bar{\beta}_t^w}\norm{\vb{w}_t - \sqrt{\bar{\alpha}_t^w}\vb{w}_0(\lambda)}^2 \\
	&= \frac{1}{\bar{\beta}_t^w} \bre{\vb{w}_t - \sqrt{\bar{\alpha}_t^w}\vb{w}_0(\lambda)}^T \pdv{\lambda} \bre{\sqrt{\bar{\alpha}_t^w}\vb{w}_0(\lambda)} \\
	&= \frac{\sqrt{\bar{\alpha}_t^w}}{\bar{\beta}_t^w} \bre{\vb{w}_t - \sqrt{\bar{\alpha}_t^w}\vb{w}_0(\lambda)}^T \pdv{\lambda}\vb{w}_0(\lambda),
\end{align} 
with evaluating the derivative of \cref{eq:app_cont_snr_forward2}
\begin{equation}
	\pdv{\lambda}\vb{w}_0(\lambda) = -\pi\sin(\lambda\pi)\vb{v}_1 + \pi\cos(\lambda\pi)\vb{v}_2.
\end{equation}
Using the relation
\begin{equation}
	\Expec{p(\vb{w}_t|\lambda)}{\bre{\vb{w}_t - \sqrt{\bar{\alpha}_t^w}\vb{w}_0(\lambda)}^T \bre{\vb{w}_t - \sqrt{\bar{\alpha}_t^w}\vb{w}_0(\lambda)}} = \Var{}{\vb{w}_t} = \bar{\beta}_t^w,
\end{equation}
we now can calculate the Fisher information directly
\begin{align}
	\mathcal{I}(\lambda) &= \int s(\lambda, \vb{w}_t)^2\; p(\vb{w}_t|\lambda) \dd{\vb{w}_t} \\
	&= \frac{\bar{\alpha}_t^w}{(\bar{\beta}_t^w)^2}\Expec{p(\vb{w}_t|\lambda)}{\br{\bre{\vb{w}_t - \sqrt{\bar{\alpha}_t^w}\vb{w}_0(\lambda)}^T \bre{-\pi\sin(\lambda\pi)\vb{v}_1 + \pi\cos(\lambda\pi)\vb{v}_2}}^2} \\
	&= \frac{\bar{\alpha}_t^w}{(\bar{\beta}_t^w)^2}\pi^2 \norm{\sin(\lambda\pi)\vb{v}_1 - \cos(\lambda\pi)\vb{v}_2}^2\; \Var{}{\vb{w}_t} \\
	&= \frac{\bar{\alpha}_t^w}{\bar{\beta}_t^w}\pi^2 \norm{\sin(\lambda\pi)\vb{v}_1 - \cos(\lambda\pi)\vb{v}_2}^2 \\
	&= \frac{\bar{\alpha}_t^w}{\bar{\beta}_t^w}\pi^2 \br{\sin(\lambda\pi)^2 \norm{\vb{v}_1}^2 + \cos(\lambda\pi)^2 \norm{\vb{v}_2}^2 - 2 \sin(\lambda\pi)\cos(\lambda\pi)\braket{\vb{v}_1}{\vb{v}_2}} \\
	&= \frac{\bar{\alpha}_t^w}{\bar{\beta}_t^w} \pi^2 d_w\\
	&= \mathrm{SNR}_w(t) \cdot \pi^2 d_w,
\end{align}
where we used our orthogonal unit-variance normalized basis for $\vb{v}_1$ and $\vb{v}_2$ (see \cref{sec:app_cont_emb}). Given the previous, we define
\begin{equation}
	\mathrm{SNR}_\lambda(t) \coloneq \mathrm{SNR}_w(t) \cdot \pi^2 d_w
\end{equation}
Therefore, for velocity prediction (see \cref{sec:app_v_weight}), we set the weighting to
\begin{align}
	\omega_w(t) 
	&\coloneq (1-\bar{\alpha}_{t}^{w}) \cdot\mathrm{sigmoid}[\log(\mathrm{SNR}_\lambda(t))] \nonumber\\
	&= (1-\bar{\alpha}_{t}^{w}) \cdot\mathrm{sigmoid}[\log(\mathrm{SNR}_w(t)) - b], \label{eq:app_w_w}
\end{align}
where $b = -\log(\pi^2 d_w)\approx -3.39$ with dimension $d_w=3$. 

\subsection{Learned noise schedule for discrete tokens}\label{sec:app_learned_schedule}

In this section, we present the details on the method used to learn the appropriate noise schedule for the discrete mode based on the given token embeddings.

\paragraph{Token mixing in continuous-state Gaussian diffusion}
The problem of too-weak-noise schedulers for diffusing discrete token embeddings is visualized in \cref{fig:DM_schedule}a, e.g. for the cosine schedule~\cite{nichol2021improved}, typically used for images. As explained in \cref{sec:embedding}, we aim to learn a noise schedule with a desired average Hamming distance profile, such that the time of trivial denoising is minimized. Note that the Hamming distance for a single token can only be either 0 (not flipped) or 1 (flipped). Instead, when averaging w.r.t. to a sampling process, we can use the average Hamming distance, which is effectively the probability of a token flipping into any other token.

Hence, instead of working directly with the Hamming distance, we define instead an analogous of its averaged version, namely the probability $p_\text{flip}(t)$ of a token initially belonging to class $i$ being decoded as any of the other $N$ classes, i.e. $j\neq i$, at time $t$:
\begin{align} \label{eq:app_p_flip}
	p_\text{flip}(t) 
	&= 1 - \Expec{\vb{h}_t^{(i)}\sim q(\vb{h}_t^{(i)}|\vb{h}_0^{(i)})}{\mathrm{softmax}_j\br{\frac{1}{\tau}\braket{\vb{h}_0^{(j)}}{\vb{h}_t^{(i)}}}}_i \nonumber\\
	&= 1 - \Expec{\vb{h}_t^{(i)}\sim q(\vb{h}_t^{(i)}|\vb{h}_0^{(i)})}
	{\frac{\exp(\frac{1}{\tau}\braket{\vb{h}_0^{(i)}}{\vb{h}_t^{(i)}})}{\sum_{j}\frac{1}{\tau}\braket{\vb{h}_0^{(j)}}{\vb{h}_t^{(i)}}}},
\end{align}
where $\tau>0$ is a temperature, which we generally set to $\tau=1/\sqrt{d_h}$ for a token embedding dimension of $d_h$. Analogous to the average Hamming distance, this probability $p_\text{flip}(t)$ is upper bounded by $1-1/N$, i.e., when $\vb{h}_t^{(i)}$ is sampled from the uniform distribution (see $t=1$ at \cref{fig:DM_schedule}a).

\paragraph{Learned discrete schedule}
In order to match $p_\text{flip}(t)$ to a desired Hamming distance target $f_\text{target}(t)$, we optimize the noise schedule $\bar{\alpha}_{t}^h$ appearing in \cref{eq:app_p_flip} by minimizing
\begin{equation}\label{eq:app_sched_loss}
	\mathcal{L}_{\text{discrete-schedule}} = \sum_{i=0}^{N-1}
	\Expec{
		t\sim\mathcal{U}\br{0,1},\,
		\vb{h}_t^{(i)}\sim q(\vb{h}_t^{(i)}|\vb{h}_0^{(i)})
	}
	{
		\norm{p_\text{flip}(t)-f_\text{target}(t)}^2
	},
\end{equation}
where we use
\begin{equation}
	p(\vb{h}_t^{(i)}|\vb{h}_0^{(i)}) = \mathcal{N}(\vb{h}_t; \sqrt{\bar{\alpha}_{t}^h}\; \vb{h}_0^{(i)}, (1-\bar{\alpha}_{t}^h) \vb{I}).
\end{equation} 
In practice, we found it easier to optimize $\bar{\alpha}_{t}^h$ indirectly by parameterizing it with $\alpha_{t}^h$, using the cumulative product relation $\bar{\alpha}_{t(i)}^h=\prod_{k=0}^{i}\alpha_{t(k)}^h$ (see \cref{eq:app_markov_transitions1} and \cref{eq:app_markov_transitions2}) and treating $\alpha_{t(k)}^h$ as learnable parameters. Before optimization, we initialize the schedule values $\alpha_{t(k)}^h$ with the cosine schedule~\cite{nichol2021improved}.

\paragraph{Visualization of different schedules}
In \cref{fig:DM_schedule}a we show the average Hamming distance of different optimized schedules according to \cref{eq:app_sched_loss}. The curves plotted correspond to the following noise schedules:
\begin{enumerate}[(i)]
	\item Linear: $f_\text{lin}(t) = (1-1/N)\, t $
	\item Sinus: $f_\text{sin}(t) = (1-1/N)\, \sin(t \pi/2) $
	\item Sinus squared: $f_\text{sin2}(t) = (1-1/N)\, \sin(t \pi/2)^2$
\end{enumerate}
In \cref{sec:experiments}, we use for all experiments a learned discrete noise schedule for the linear target (see \cref{fig:DM_schedule}a and c).

\paragraph{Uniform discrete-state diffusion}
As shown in \cref{sec:learned_sched}, we use the duality between uniform discrete-state diffusion and continuous-state Gaussian diffusion~\cite{sahoo2025diffusionduality} to write the probability of finding a decoded, one-hot encoded token $\vb{k}_t\in\reals^N$ for a discrete schedule $a_t$ as
\begin{equation}\label{eq:app_cat_DM}
	p(\vb{k}_t|\vb{k}) = \mathrm{Cat}(\vb{k}_t; a_t \vb{k} + (1-a_t)\vb{I}/N).
\end{equation}
From this, we define the SNR of the discrete mode as the fraction of the original amplitude to the one of the uniform distribution 
\begin{equation}\label{eq:app_discrete_snr}
	\mathrm{SNR}_{\text{discrete}}(t) \coloneq \frac{a_t}{1 - a_t}.
\end{equation}
Just as the continuous SNR (see \cref{eq:gaussian_dm}), we have the same boundary behavior of no noise at small diffusion times ($\lim_{t\rightarrow0}\mathrm{SNR}_{\text{discrete}}(t)\rightarrow\infty$) to full noise at time $t=1$ ($\mathrm{SNR}_{\text{discrete}}(t=1)=0$). Here, full noise means a uniform distribution across the classes.

Similar to the continuous mode, we use the sigmoid log-SNR weighting of \cref{eq:app_sigmoid_weight} (see \cref{sec:multidiff}). In order to account for the discrete nature of the embeddings, which embody high correlations across the embedding dimensions, we adjust the normal $\mathrm{SNR}_h(t)$ (\cref{sec:multidiff}) to the discrete one $\mathrm{SNR}_{\text{discrete}}(t)$ of \cref{eq:app_discrete_snr}.

To determine the appropriate weight $\omega_h(t)$ in \cref{eq:loss}, we relate the Gaussian flip probabilities $p_\text{flip}(t)$ (\cref{eq:app_p_flip}) to the discrete schedule $a_t$ of \cref{eq:app_cat_DM}. As all tokens are equivalently encoded, it suffices to look at only one token $i$. Hence, following \cref{eq:app_cat_DM}, we write the probability of finding the token not to be flipped as
\begin{equation}
	\brek{a_t \vb{k} + (1-a_t)\vb{I}/N}_i = a_t + (1-a_t) \frac{1}{N}	
	\stackrel{!}{=} p_\text{not flipped}(t) = 1 - p_\text{flip}(t),
\end{equation}
resulting in the claimed relation of \cref{sec:learned_sched}
\begin{equation}
	a_t = 1 - \frac{p_\text{flip}(t)}{1 - 1/N} = 1 - \frac{p_\text{flip}(t)}{p_\text{flip}(1)}.
\end{equation}
Therefore, assuming a well optimized learned schedule with Hamming target $f_\text{target}(t)$ (see \cref{eq:app_sched_loss}), we can rewrite the discrete SNR as
\begin{align}
	\mathrm{SNR}_{\text{discrete}}(t) 
	&= \frac{a_t}{1 - a_t}
	\nonumber\\
	&= \frac{1 -\bre{p_\text{flip}(t)/p_\text{flip}(1)}}{p_\text{flip}(t)/p_\text{flip}(1)} 
	\nonumber\\
	&= \frac{p_\text{flip}(1) - p_\text{flip}(t)}{p_\text{flip}(t)} 
	\nonumber\\
	&\approx \frac{f_\text{target}(1) - f_\text{target}(t)}{f_\text{target}(t)}.
\end{align}
Finally, for the velocity prediction (see \cref{sec:app_v_weight}), we set the weighting to
\begin{align}
	\omega_h(t)
	&\coloneq (1-\bar{\alpha}_{t}^{h}) \cdot\mathrm{sigmoid}[\log(\mathrm{SNR}_{\text{discrete}}(t))]
	\nonumber\\
	&= (1-\bar{\alpha}_{t}^{h}) \cdot\mathrm{sigmoid}\bre{\log(\frac{f_\text{target}(1) - f_\text{target}(t)}{f_\text{target}(t)})}. \label{eq:app_w_h}
\end{align}

\section{Circuit encoding and embedding}\label{sec:app_emb}

Starting from a quantum circuit $\vb{x}$, we first tokenize the gates into an integer-matrix representation (see \cref{fig:app_DM_overview}), following the method of Ref.~\cite{furrutter2024quantum}. For this, each gate is assigned to a distinct token. Moreover, a sign is added to specify the connection type, e.g. control connections are implemented as negative tokens and target connections as positive ones. The resulting matrix has then dimension equal to the number of qubits $n$ times the number of gates $t$, i.e. $\mathrm{tokenize}(\vb{x})\in\integer^{n \cross t}$, which is the format we use to store the circuits of the dataset (see \cref{sec:app_dataset}).

In this work, extending from the previous representation, we consider additional continuous parameters for the parameterized gates. We store them alongside the token matrix as a one dimensional array matching the sequence length $t$ (see \cref{fig:app_DM_overview}), i.e. $\mathrm{parameters}(\vb{x})\in\reals^{t}$. Importantly, the values of this array for non-parametrized gates are set to zero.

After tokenization, in order to diffuse a circuit $\vb{x}$ via the forward process defined in \cref{eq:forward_multimodial}, we embed it into a continuous representation $[\vb{h}_0, \vb{w}_0]$, which we discuss in the following  \cref{sec:app_discr_emb} and \cref{sec:app_cont_emb}.

\subsection{Discrete token embedding}\label{sec:app_discr_emb}

As described in \cref{sec:embedding}, we implement the token embeddings as $d_h$-dimensional entries $\vb{h}_0^{(i)}\in\reals^{d_h}$ of a look-up table, where $i  = {0, \dots, N-1}$ indexes the $N$ different classes. To ensure that all embeddings are equidistant and undergo uniform mixing throughout the diffusion process, we construct them as an orthogonal basis of $\reals^{d_h}$.

In addition, we set further constraints on our embeddings, $\forall i\in \{0, \dots, N-1 \}$
\begin{align}
	\expecOp[{\vb{h}_0^{(i)}}]  &= 0   \quad\text{and} \label{eq:app_emb_restrictions1}\\
	\varOp[{\vb{h}_0^{(i)}}] &= 1,  \label{eq:app_emb_restrictions2}
\end{align}
where the expectation and variance are taken over the vector elements. 

The first point \cref{eq:app_emb_restrictions1} addresses the known \textit{average brightness issue} of DMs~\cite{zhang2024tackling, lin2024common}, where they conserve the same mean as that of the starting noise latent (at $t=1$) during the whole ancestral sampling process. 
This property is attributed to the fact that the low frequency information of the original data $\vb{x}$ survives the forward diffusion process longer than high frequency details, enabling the DM to use the low frequencies as a help to lower the loss. For instance, as images have rarely zero mean, the latent initialization of $\vb*{\epsilon}\sim\mathcal{N}(\vb{0}, \vb{I})$ at $t=1$ biases the DM towards very specific images, i.e. those with zero average brightness. For circuit generation, we found the low frequency information results in the circuit length to be easily predicable during training, stemming from the fact that the padding token count offsets the mean of a whole circuit embedding linearly, when not guaranteeing property \cref{eq:app_emb_restrictions1}. The second point \cref{eq:app_emb_restrictions2} accounts for a variance preserving diffusion process~\cite{karras2022elucidating}, where we require normalized data with unit variance. 

Remarkably, by considering the constraints \cref{eq:app_emb_restrictions1} and \cref{eq:app_emb_restrictions2} for single token embeddings, all possible circuits are always embedded with zero-mean and unit-variance by construction. Note, the orthogonality together with the constraints (\cref{eq:app_emb_restrictions1} and \cref{eq:app_emb_restrictions2}) allow a maximum of $N=d_h-1$ classes, as the zero-sum reduces one dimension. For the gate set of \cref{sec:experiments} (i.e. tokens $\brek{\mathrm{empty}, \mathrm{h},\pm\mathrm{cx},\pm\mathrm{ccx},\mathrm{swap},\mathrm{rx},\mathrm{ry},\mathrm{rz},\mathrm{cp}, \mathrm{padding}}$) we have $N=12$ connection types in total. Therefore, we use the embedding dimension of $d_h=13$.

\subsection{Continuous parameter embedding}\label{sec:app_cont_emb}

The normalized continuous parameters $\lambda\in\bre{-1,1}$ are encoded as
\begin{equation}
	\vb{w}_0 = \cos(\lambda\pi) \vb{v}_1 + \sin(\lambda\pi) \vb{v}_2,
\end{equation}
where both $\vb{v}_1, \vb{v}_2 \in \reals^{d_w}$  fulfill the conditions \cref{eq:app_emb_restrictions1} and \cref{eq:app_emb_restrictions2}. Further, we set $\vb{v}_1$ and $\vb{v}_2$ to be orthogonal, fixing $d_w=3$. Given a noisy embedding $\vb{w}_t$, we can decode the parameter using the estimator
\begin{equation}
	\hat{\lambda}_t = \frac{1}{\pi} \mathrm{arctanh2}\br{\frac{\braket{\vb{v}_2}{\vb{w}_t}}{\braket{\vb{v}_1}{\vb{w}_t}}}.
\end{equation}
To analyze the diffusion behavior induced by this encoding, we define a distance which accounts for the periodicity of the parameters
\begin{equation}
	\mathrm{CircularLoss}(t) \coloneq 1 - \cos((\lambda-\hat{\lambda}_t)\pi).
\end{equation}
In \cref{fig:DM_schedule}b we show the $\mathrm{CircularLoss}(t)$ of different noise schedules, which are defined as follows:
\begin{enumerate}[(i)]
	\item Linear: The linear beta schedule of DDPM~\cite{ho2020denoising}: $\beta_t^w=(1-t)\beta_0^w + t \beta_1^w$, with $\beta_0^w=10^{-4}$ and $\beta_1^w=0.02$.
	\item Cosine: The cosine schedule~\cite{nichol2021improved}: $\bar{\alpha}_t^w=\cos(t \pi/ 2)$.
	\item Cosine squared: The squared version of the cosine schedule: $\bar{\alpha}_t^w=\cos(t \pi/ 2)^2$.
\end{enumerate}
In \cref{sec:experiments}, we use for all experiments the cosine squared noise schedule for the continuous mode (see \cref{fig:DM_schedule}b and c). 

\section{Additional experiments}\label{sec:app_add_experiments}

\subsection{Comparison to existing methods: QFT} \label{sec:app_comparison_compilers}

Here we compare the compilers of \cref{sec:comparison} on a single unitary -- the Quantum Fourier transform (QFT) unitary. We report in \cref{tab:compiler_comparison_qft} the synthesis results and the known textbook solution~\cite{coppersmith2002approximate}. Here, we use default compiler settings, without restricting their runtimes. For genQC2, we generate 2048 circuits and select the circuit with minimum infidelity, as mentioned \cref{sec:discovery}. For the ansatz improvements, we use the same settings as in \cref{sec:app_tree_search}. We find that our model and the ansatz fixes are capable of finding the exact textbook circuit, only missing some numeric precision in the optimized angles.

\begin{table}[!ht]
	\caption{\textbf{Synthesis methods comparison: QFT.} Infidelity and gate count of found circuits for the QFT unitary.}
	\label{tab:compiler_comparison_qft}
	\centering
	\begin{tabular}{lcrr}
		\toprule
		Method & \# qubits & Infidelity & Gate count \\

		\midrule  
		Textbook circuit                         & 3 & ~ &  7 \\ 
		(Ref. \cite{coppersmith2002approximate}) & 4 & ~ & 12 \\ 
		~ 					                     & 5 & ~ & 17 \\ 

        \midrule
        genQC2 (ours)  & 3 & $1.4 \cdot 10^{-4}$ &  7 \\ 
		~ 			   & 4 & $3.8 \cdot 10^{-3}$ & 12 \\ 
		~ 			   & 5 & $7.8 \cdot 10^{-2}$ & 15 \\ 
        
    	\midrule		
        + ansatz fixes          & 3 & $1.3 \cdot 10^{-7}$ &  7 \\ 
        (\cref{sec:gen_ansatz}) & 4 & $9.2 \cdot 10^{-7}$ & 12 \\ 
        ~ 				        & 5 & $3.4 \cdot 10^{-9}$ & 17 \\ 

		\midrule  
		QSD                                          & 3 & $8.9 \cdot 10^{-16}$ &   61 \\ 
		(Refs.~\cite{shende2006qsd, krol2024beyond}) & 4 & $3.6 \cdot 10^{-15}$ &  281 \\ 
		~ 					                         & 5 & $3.0 \cdot 10^{-12}$ & 1540 \\ 

        \midrule   
		AQC                          & 3 & $2.0 \cdot 10^{-9}$ &   79 \\ 
		(Ref.~\cite{madden2022best}) & 4 & $5.9 \cdot 10^{-5}$ &  317 \\ 
		~ 	                         & 5 & $1.0 \cdot 10^{-4}$ & 1275 \\ 
        
        \midrule    
		LEAP                         & 3 & $8.9 \cdot 10^{-16}$ &  70 \\ 
		(Ref.~\cite{smith2023LEAP}) & 4 & $6.7 \cdot 10^{-16}$ & 130 \\ 
		~ 					          & 5 & $1.2 \cdot 10^{-14}$ & 212 \\ 
 
		\bottomrule
	\end{tabular}
\end{table}

\subsection{Comparison to existing methods: restricted gate count} \label{sec:app_comparison_compilers_restircted}

In this section, we want to compare our method to the approximate compilers of \cref{tab:compiler_comparison}, i.e. AQC and LEAP, with the new addition of QFAST~\cite{younis2021qfast}. Importantly, here we limit the amount of multi-qubit gates the comilers are allowed to use and report the results in \cref{tab:compiler_comparison_restricted}. As in \cref{sec:comparison}, to speed up the simulations, we limit the number of circuit samples per unitary  and timeout compilations which take more than one hour for one unitary. For LEAP and QFAST, we sample for 3 qubits 4 circuits per unitary, for 4 qubit 2 circuits per unitary, and for 5 qubits only one. Further, for AQC we sample for 3, 4 and 5 qubits each 128 circuits per unitary.

\begin{table*}
	\caption{\textbf{Synthesis methods comparison with restricted gate counts.} Reported values are averages over random 480 unitaries (from 2 to 16 gates) and 128 circuits sampled for each unitary. Values in parenthesis are the corresponding standard deviation, showing the variability of the metrics. We limit the number of CNOT layers for AQC, for LEAP the number of multi-qubit gates, and QFAST the number of 2-qubit gates. The lower bound of AQC refers to the theoretical lower bound of required CNOT gates~\cite{madden2022best}. (*)~Some sample restrictions apply, as detailed in \cref{sec:comparison} and \cref{sec:app_comparison_compilers_restircted}. (**)~Not reported due to lower sample count.
    }
	\label{tab:compiler_comparison_restricted}
	\centering
	\begin{tabular}{
			p{2.25cm}
			c
            >{\raggedleft\arraybackslash}p{2.7cm}
			>{\raggedleft\arraybackslash}p{3.65cm}
			>{\raggedleft\arraybackslash}p{1.7cm}
            >{\raggedleft\arraybackslash}p{2.2cm}
			>{\raggedleft\arraybackslash}p{2cm}
		}
		\toprule
        Method & \#~qubits & {Limit \# of \par multi-qubit \par gates} & Avg. minimum\par infidelity & Avg. gate count & Avg. runtime per sample (seconds) & Avg. distinct circuits per unitary \\ 

        \midrule \midrule	
		genQC2 (ours)		& 3 & NA & $0.09\;(0.20)$ & $8\;(4)$ & $0.09\;(0.00)$ & $78\;(45)$ \\ 
		~ 					& 4 & NA & $0.10\;(0.20)$ & $8\;(4)$ & $0.09\;(0.00)$ & $74\;(49)$ \\ 
		~ 					& 5 & NA & $0.09\;(0.17)$ & $8\;(4)$ & $0.09\;(0.00)$ & $71\;(50)$ \\ 
        
        \midrule		
		+ ansatz fixes & 3 & NA & $0.008\;(0.037)$ & $ 9\;(5)$ & $190\;(390)$ & NA \\ 
		  (\cref{sec:gen_ansatz}) & 4 & NA & $0.015\;(0.062)$ & $ 9\;(5)$ & $330\;(570)$ & NA \\ 
		  ~ & 5 & NA & $0.015\;(0.071)$ & $ 9\;(5)$ & $480\;(880)$ & NA \\ 
    
        \midrule \midrule
		AQC*                         & 3 & 1 & $0.67\;(0.20)$ & $13\;(1)$   & $0.01\;(0.00)$  & $7\;(8)$ \\  
		(Ref.~\cite{madden2022best}) & ~ & 2 & $0.63\;(0.17)$ & $18\;(1)$   & $0.02\;(0.01)$  & $12\;(13)$ \\  
        ~ & ~ &                4 & $0.31\;(0.23)$ & $28\;(1)$   & $0.03\;(0.01)$  & $32\;(26)$ \\ 
        ~ & ~ & (low. bound)~~14 & $1.0\cdot 10^{-9}\;(3.8\cdot 10^{-9})$ & $79\;(0)$   & $0.3\;(0.2)$  & $21\;(12)$ \\ 
        ~ & ~ &               16 & $6.1\cdot 10^{-10}\;(7.5\cdot 10^{-10})$ & $89\;(0)$  & $0.3\;(0.1)$  & $10\;(7)$ \\ 
        \midrule  
		~ & 4 &  1 & $0.80\;(0.15)$ & $16\;(1)$  & $0.02\;(0.01)$  & $10\;(10)$ \\ 
		~ & ~ &  2 & $0.81\;(0.13)$ & $21\;(1)$  & $0.03\;(0.01)$  & $13\;(15)$ \\ 
        ~ & ~ &  4 & $0.73\;(0.15)$ & $31\;(1)$  & $0.04\;(0.02)$  & $25\;(24)$ \\ 
        ~ & ~ & 16 & $0.037\;(0.081)$ & $88\;(2)$ & $0.3\;(0.2)$  & $97\;(33)$ \\ 
        ~ & ~ & (low. bound)~~61 & $4.3\cdot 10^{-5}\;(5.4\cdot 10^{-5})$ & $317\;(0)$  & $5.2\;(0.1)$  & ~$^{**}$ \\ 
        \midrule 
		~ & 5 &  1 & $0.83\;(0.18)$ & $19\;(2)$ & $0.04\;(0.01)$ & $12\;(12)$ \\ 
		~ & ~ &  2 & $0.88\;(0.10)$ & $24\;(1)$ & $0.05\;(0.02)$ & $13\;(14)$ \\
        ~ & ~ &  4 & $0.90\;(0.06)$ & $34\;(1)$ & $0.07\;(0.03)$ & $18\;(17)$ \\ 
        ~ & ~ & 16 & $0.28\;(0.26)$ & $91\;(2)$ & $0.3\;(0.2)$ & $104\;(27)$ \\ 
        ~ & ~ & (low. bound)~~252 & $1.0\cdot 10^{-4}\;(0.3\cdot 10^{-4})$ & $1275\;(0)$ & $27.7\;(0.1)$ & ~$^{**}$ \\ 

        \midrule \midrule
        LEAP*                       & 3 & 1 & $0.44\;(0.27)$ & $27\;(3)$ & $1.2\;(0.1)$ & ~$^{**}$ \\
		(Ref.~\cite{smith2023LEAP}) & ~ & 2 & $0.22\;(0.23)$ & $38\;(7)$ & $1.9\;(0.5)$ & ~$^{**}$ \\
        ~ & ~ & 4        & $0.028\;(0.065)$            & $53\;(17)$ & $3.0\;(9.5)$ & ~$^{**}$\\
        ~ & ~ & 16       & $1.5\cdot 10^{-3}\;(3.8\cdot 10^{-3})$ & $60\;(26)$ & $1.6\;(0.6)$ & ~$^{**}$  \\ 
        ~ & ~ & no limit & $0.4\cdot 10^{-3}\;(1.9\cdot 10^{-3})$ & $61\;(26)$ & $0.8\;(0.6)$ & $41\;(39)$ \\ 
        \midrule 
		~ & 4 & 1 & $0.56\;(0.30)$ & $32\;(3)$ & $1.6\;(0.2)$ & ~$^{**}$ \\
        ~ & ~ & 2 & $0.36\;(0.30)$ & $43\;(6)$ & $5.3\;(3.0)$ & ~$^{**}$ \\
        ~ & ~ & 4        & $0.14\;(0.21)$              & $59\;(16)$ & $6\;(21)$ & ~$^{**}$ \\
        ~ & ~ & 16       & $2.5\cdot 10^{-3}\;(6.7\cdot 10^{-3})$ & $86\;(50)$ & $25\;(85)$ & ~$^{**}$ \\
        ~ & ~ & no limit & $1.2\cdot 10^{-3}\;(3.2\cdot 10^{-3})$ & $85\;(50)$ & $40\;(250)$ & ~$^{**}$ \\
        \midrule 
		~ & 5 & 1 & $0.59\;(0.34)$ & $36\;(3)$ & $3.3\;(1.1)$ & ~$^{**}$ \\ 
        ~ & ~ & 2 & $0.42\;(0.35)$ & $47\;(7)$ & $50\;(40)$   & ~$^{**}$ \\ 
        ~ & ~ & 4        & $0.21\;(0.29)$              & $63\;(17)$ & $45\;(77)$ & ~$^{**}$\\
        ~ & ~ & 16       & $0.012\;(0.041)$            & $95\;(61)$ & $240\;(600)$ & ~$^{**}$ \\ 
        ~ & ~ & no limit & $2.7\cdot 10^{-3}\;(5.3\cdot 10^{-3})$ & $92\;(57)$ & $170\;(500)$ & ~$^{**}$ \\ 
        
        \midrule \midrule
		QFAST*                        & 3 & 1 & $0.65\;(0.24)$ & $20\;(2)$ & $1.6\;(0.1)$ & ~$^{**}$ \\ 
		(Ref.~\cite{younis2021qfast}) & ~ & 4 & $0.25\;(0.21)$ & $47\;(10)$ & $1.7\;(0.2)$ & ~$^{**}$ \\ 
		~ & ~ & 16 & $0.046\;(0.070)$ & $80\;(30)$ & $2.0\;(0.6)$ & ~$^{**}$ \\
        ~ & ~ & 36 & $2.7\cdot 10^{-3}\;(5.2\cdot 10^{-3})$ & $99\;(45)$ & $2.4\;(1.2)$ & ~$^{**}$ \\
        \midrule 
		~ & 4 &  1 & $0.84\;(0.20)$ & $20\;(4)$  &  $1.9\;(0.2)$ & ~$^{**}$ \\ 
		~ & ~ &  4 & $0.65\;(0.27)$ & $45\;(10)$ &  $2.3\;(0.3)$ & ~$^{**}$ \\ 
        ~ & ~ & 16 & $0.38\;(0.27)$ & $80\;(23)$ &  $2.9\;(0.8)$ & ~$^{**}$ \\ 
        ~ & ~ & 36 & $0.23\;(0.22)$ & $115\;(37)$ & $4.0\;(1.7)$ & ~$^{**}$ \\ 
        \midrule 
		~ & 5 &  1 & $0.92\;(0.15)$ & $19\;(4)$ & $2.7\;(0.5)$ & ~$^{**}$ \\ 
        ~ & ~ &  4 & $0.83\;(0.22)$ & $44\;(11)$ & $3.5\;(0.8)$ & ~$^{**}$ \\ 
        ~ & ~ & 16 & $0.65\;(0.30)$ & $80\;(20)$ & $6.5\;(2.7)$ & ~$^{**}$ \\ 
        ~ & ~ & 36 & $0.54\;(0.32)$ & $115\;(33)$ & $13.0\;(8.1)$ & ~$^{**}$ \\ 

		\bottomrule
	\end{tabular}
\end{table*}

\subsection{Comparison to genQC1} \label{sec:app_genqc_1vs2}

We compare the method presented in this paper (genQC2) to its predecesor genQC1~\cite{furrutter2024quantum}. To summarize, the major upgrades from genQC1 are:
\begin{enumerate}
	\item New multi-modal diffusion with separated discrete and continuous modes, allowing native generation of parameterized gates jointly with their continuous control parameters, without post-gradient-based parameter optimization. (\cref{sec:app_multimode_details})
	\item Completely new transformed-based architecture (CirDiT; \cref{sec:app_model_arch}).
	\item Pre-training of a custom Unitary-CLIP (\cref{sec:app_unitary_clip}).
	\item Scaling to 5 qubits with up to 32 gates (\cref{sec:app_dataset}).
	\item General improvements in learning: learned noise schedulers to better account for the variance of both modes, adjusted loss weighting with mode specific signal-to-noise ratios, improved discrete embedding choice (zero mean, unit variance) (\cref{sec:app_multimode_details} and \cref{sec:app_emb}).
\end{enumerate}

\begin{figure}
	\centering
	\includegraphics[width=0.85\columnwidth]{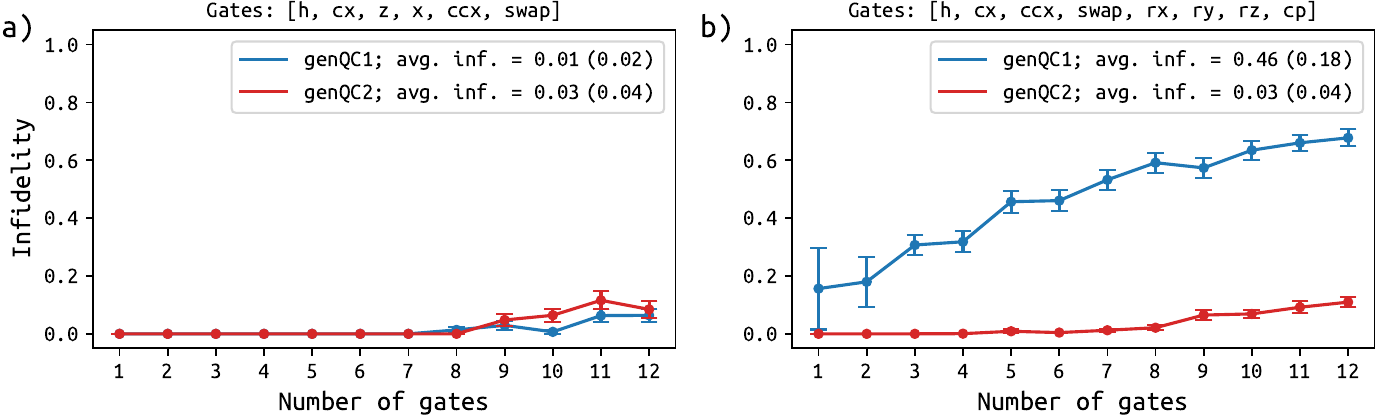} 
	\caption{\textbf{Comparison genQC 1 vs 2.}
		Average minimum infidelities vs number of gates of the target unitaries. Error bars are the error of the means. Values in parenthesis of the legend are the standard deviation, showing variability of the average infidelities. The target unitaries are over 3 qubits, up to a maximum number of 12 gates, and are sampled with: \textbf{a)} the discrete gate set of genQC1; \textbf{b)} the discrete-continuous gate set of genQC2.
	}
	\label{fig:comparison_genqc_1vs2}
\end{figure}

To directly compare the compilation accuracy of genQC1 and genQC2, we have to restrict the unitaries to the circuit limitations of genQC1, which are only 3 qubits and a maximum number of 12 discrete gates. In \cref{fig:comparison_genqc_1vs2}, we show a comparison of the infidelities on the gate sets each of the two models were trained on, using, respectively, 725 and 650 unitaries. We sample 128 circuits per unitary and record the minimum infidelity. We note that the two models compile into their respective gate sets, while the target unitary gate set varies. In particular, genQC1 was trained on discrete circuits while genQC2 also consider continuous gates. Importantly, we observe that genQC2 is reaching a similar performance as genQC1 on the discrete gate set, without being trained explicitly for it. Contrary, genQC1 is not able to compile unitaries consisting of parameterized gates with low infidelity.

\subsection{Post-optimization of continuous gates} \label{sec:app_classic_opt}

We analyze now the accuracy of the continuous parameters of the generated circuits, i.e. the performance of the continuous mode. For this, we explore how much a given circuit can be improved using a standard gradient-based optimization method. We then compare these optimized results with those obtained from optimizing the same circuits whose continuous parameters were uniformly sampled.

\begin{figure}
	\centering
	\includegraphics[width=0.90\columnwidth]{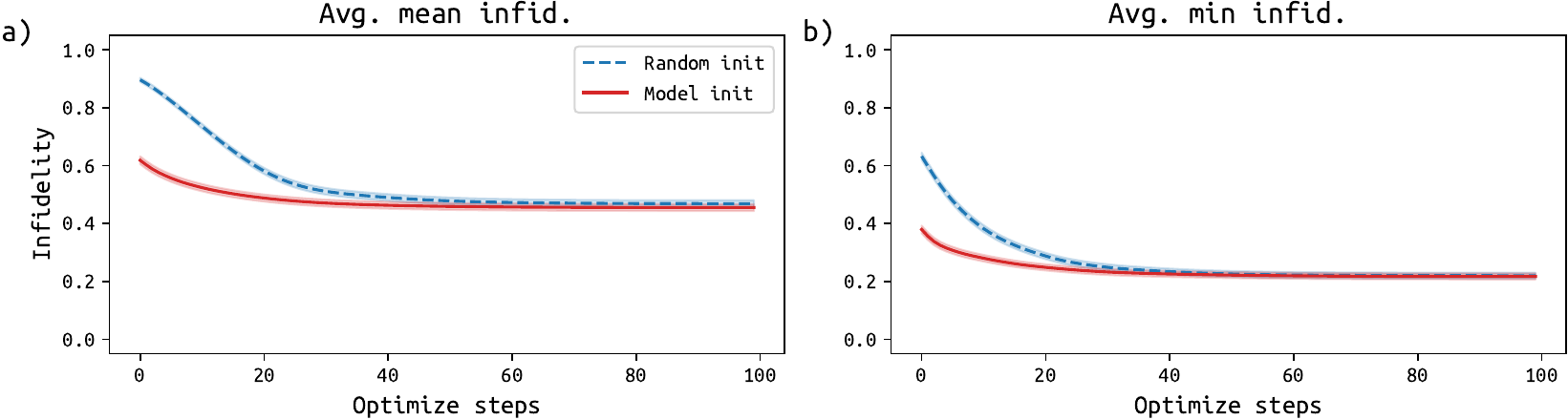} 
	\caption{\textbf{Post-optimization.} Gradient-based optimization of the continuous parameters of the generated circuits, starting with the parameters predicted by the model or randomly initialized parameters. Shown are in \textbf{a)} the average mean infidelities, and in \textbf{b)} the average minimum infidelities. Shaded areas are the errors of the means. Details in \cref{sec:app_classic_opt}.
	}
	\label{fig:post_optimize}
\end{figure}

For the results shown in \cref{fig:post_optimize}, we run the optimizer on generated circuits for 3 to 5 qubits, each 256 unitaries, uniformly consisting of 2 to 32 gates. We consider 128 generated circuits per unitary. For the gradient-based optimization of the continuous parameters, we use the backpropagation method, a learning rate (lr) of $0.1$ with a cosine annealing lr-schedule and a maximum of 100 update steps. To speed up the optimization, we stop early if the infidelity does not change anymore than $10^{-12}$.

We see in \cref{fig:post_optimize} that the predicted initialization of the generated circuits has a notable lower infidelity than random starting points. We also observe a faster convergence of the generated parameters, accelerating the optimization and reducing the number of update steps required. The fact that the optimizer does not reach zero is attributed to the erroneous placement of gates in the generated ansätze. We discuss possible fixes for this in \cref{sec:gen_ansatz}.

\subsection{Tree search details} \label{sec:app_tree_search}

\begin{figure}
	\centering
	\includegraphics[width=0.6\columnwidth]{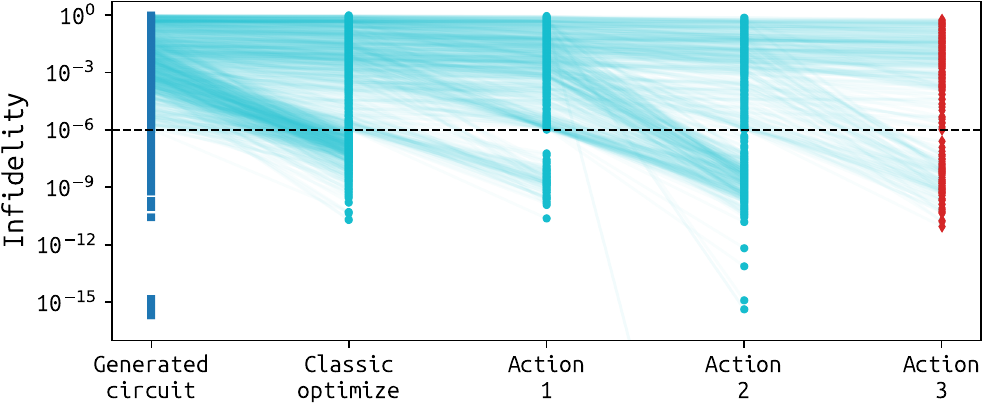}
	\caption{\textbf{Tree search paths.}
		Infidelity trajectories of all considered circuits after every action is applied. The dotted black line represents the expansion threshold, where each trajectory stops once it passed this threshold (see \cref{sec:app_tree_search}). Generated circuits and their optimizations are the same as in \cref{fig:ansatz_fixing}.
	}
	\label{fig:tree_search_paths}
\end{figure}

We provide here additional details on the tree search method presented in \cref{sec:gen_ansatz}. We run the search on generated circuits for 3 to 5 qubits, each 480 unitaries, uniformly consisting of 2 to 16 gates. Detailed results are presented in \cref{tab:compiler_comparison} and \cref{fig:ansatz_fixing}. In addition, in \cref{fig:tree_search_paths}, we show the trajectories of the circuit improvements.

We perform a top-$k$ greedy expansion policy, where we set $k=15$. For this, we only expand the $k$-lowest valued nodes, measured by the infidelities, at a given depth. The available actions are $\mathrm{delete}$, $\mathrm{insert}$ or $\mathrm{replace}$ a single gate. The gate set is the same as the model is using: $\brek{\mathrm{h},\mathrm{cx},\mathrm{ccx},\mathrm{swap},\mathrm{rx},\mathrm{ry},\mathrm{rz},\mathrm{cp}}$. A node expansion then considers all the possible combinations of these action with all possible gate placement possibilities. After each action, we additional perform a gradient-based optimization of the continuous parameters, before recording the new node infidelity values. To speed up the tree-search we stop expanding nodes which have an infidelity under a given $\epsilon$-threshold. We generally use $\epsilon=10^{-6}$, reflected in \cref{fig:ansatz_fixing}c and \cref{fig:tree_search_paths}.

For the gradient-based optimization of the continuous parameters, we use the backpropagation method, a learning rate (lr) of $0.1$ with a cosine annealing lr-schedule and a maximum of 1000 update steps. To speed up the optimization, we stop early if the infidelity does not change anymore than $10^{-8}$.

\subsubsection{What gates and angles are added?} \label{sec:app_tree_search_add_gates}

\begin{figure}
	\centering
	\includegraphics[width=0.85\columnwidth]{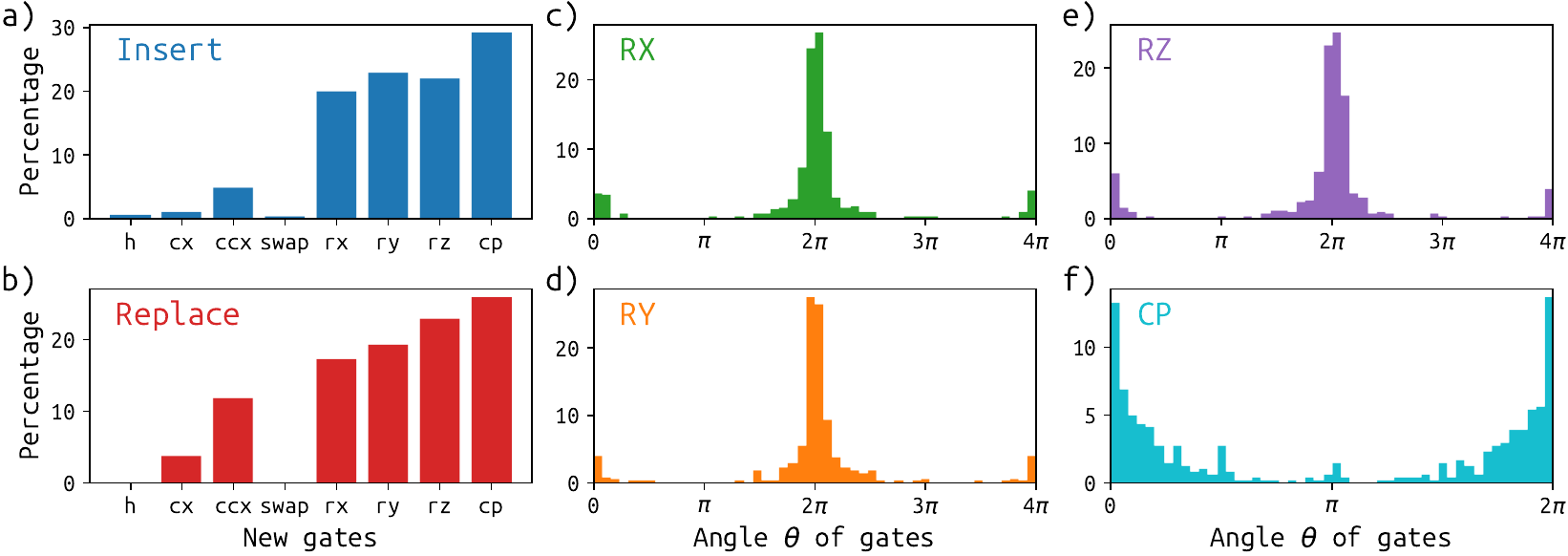}
	\caption{\textbf{Tree-search fixes.} 
        Generated circuits and their optimizations are the same as in \cref{fig:ansatz_fixing}.
		\textbf{a, b)} Distribution of new gates, added by the insert and replace actions.
        \textbf{c, d, e, f)} Angle distribution of the new gates of all actions, after the corresponding optimization.
	}
	\label{fig:tree_search_add_gates}
\end{figure}

To understand better what kind of fixes the tree-search found (see \cref{sec:gen_ansatz}), we now look at what gates are inserted and what angle distribution they have after optimization. We present in \cref{fig:tree_search_add_gates}ab the distribution of new gates which were added by the insert or replace action of the tree-search. We see most of the time a continuous gate is added, which we attribute to the fact that enlarging the ansatz, i.e. making it more expressive, is often the best action that a greedy policy can do.

Recall, the parametrized gates are defined as~\cite{cudaq}:
\begin{align}  
\mathrm{rx}(\theta) &= \begin{pmatrix} \cos(\theta/2) & -i\sin(\theta/2) \\ 
                                    -i\sin(\theta/2) & \cos(\theta/2) \end{pmatrix},  \quad
&\mathrm{ry}(\theta) &= \begin{pmatrix} \cos(\theta/2) & -\sin(\theta/2) \\ 
                                    \sin(\theta/2) & \cos(\theta/2) \end{pmatrix},  \nonumber\\
\mathrm{rz}(\theta) &= \begin{pmatrix}
\exp(-i \theta/2) &   \\
 & \exp(i \theta/2)  \\
\end{pmatrix},
&\mathrm{cp}(\theta) &= \begin{pmatrix}
1 &  &  &  \\
 & 1 &  &  \\
 &  & 1 &  \\
 &  &  & \exp(i \theta) 
\end{pmatrix}. \label{eq:app_gate_def}
\end{align} 

Note that the $\mathrm{rx}$, $\mathrm{ry}$ and $\mathrm{rz}$ gates are $4\pi$ periodic, and the $\mathrm{cp}$-gate has a period of $2\pi$. In \cref{fig:tree_search_add_gates}c-f, we show the angle distributions of all the added gates after the gradient-based optimization is done. For all of the gates, we find two main peaks, one around zero and one around $2\pi$. First, we see for $\theta=0$ that $\mathrm{rx}(0)=\mathrm{ry}(0)=\mathrm{rz}(0)=\mathrm{cp}(0)=\vb{I}$, and secondly for $\theta=2\pi$ we have  $\mathrm{rx}(2\pi)=\mathrm{ry}(2\pi)=\mathrm{rz}(2\pi)=-\vb{I}$. Notably, both $\pm\vb{I}$ have no effect on the infidelity of a circuit, as they are just global phases (see \cref{eq:infidelity}), effectively turning off these added gates. The distribution around these peaks, in \cref{fig:tree_search_add_gates}c-f, corresponds to the actual corrections the optimizer makes to the circuit ansatz. We see they are relatively close to their peaks, making the corrections rather small perturbations around the identity.

\subsection{Ablation study}\label{sec:app_ablation}

We list here additional details on the ablation studies of \cref{sec:ablation}.

\paragraph{Discrete noise schedule and weighting}
In \cref{tab:ablation_discrete_schedule}, we fix the continuous noise schedule (i.e. cosine squared) while varying only the discrete schedule and switching on our loss weighting. Results are the averages of the models presented in \cref{fig:model_ablation_scale}a, which are trained for 250k update steps.

\begin{table}[ht!]
	\caption{\textbf{Ablation of discrete noise schedules and loss weighting.} Results are averages across 3 different model sizes and for 3 to 5 qubits, more details in \cref{sec:app_ablation} and \cref{fig:model_ablation_scale}a. Reported errors are the error of the means. Note that the reported numbers are averages across different model sizes, showing the average influence of our improvements and not a single model performance.}
	\label{tab:ablation_discrete_schedule}
	\centering
	\begin{tabular}{llcc}
		\toprule
		Discrete noise schedule & Weighting $\omega_h(t), \omega_w(t)$ & Avg. mean infid. $\downarrow$ & Avg. min infid. $\downarrow$ \\ 
		\midrule
		DDPM linear (Ref. \cite{ho2020denoising}) & constant & $0.780\pm0.004$ & $0.472\pm0.006$ \\
		Cosine (Ref. \cite{nichol2021improved}) & constant & $0.772\pm0.004$ & $0.463\pm0.006$ \\
		Linear learned (Sec. \ref{sec:learned_sched}) & constant & $0.709\pm0.005$ & $0.380\pm0.006$ \\ 
		Linear learned (Sec. \ref{sec:learned_sched}) & adjusted (Sec. \ref{sec:embedding} + \ref{sec:learned_sched}) & $\textbf{0.672}\pm0.005$ & $\textbf{0.358}\pm0.005$ \\   
		\bottomrule
	\end{tabular}
\end{table}

\paragraph{Continuous noise schedule}
In \cref{tab:ablation_continuous_schedule}, we fix the discrete schedule (i.e. learned linear) and loss weighting (i.e. adjusted) while varying only the continuous noise schedule. For this ablation, we train models with 32M parameters for 250k update steps. We find no significant difference and use for all of our experiments the cosine squared schedule.

\begin{table}[ht!]
	\caption{\textbf{Ablation of continuous noise schedules.} Results are averages for 3 to 5 qubits for a single model for each setting, more details in \cref{sec:app_ablation}. Reported errors are the error of the means.}
	\label{tab:ablation_continuous_schedule}
	\centering
	\begin{tabular}{lcc}
		\toprule
		Continuous noise schedule & Avg. mean infid. $\downarrow$ & Avg. min infid. $\downarrow$ \\ 
		\midrule
		DDPM linear (Ref. \cite{ho2020denoising})    & $0.735\pm0.008$ & $0.418\pm0.010$ \\
		Cosine (Ref. \cite{nichol2021improved})      & $0.733\pm0.008$ & $0.410\pm0.010$ \\
		Cosine squared (see \cref{sec:app_cont_emb}) & $0.733\pm0.008$ & $0.407\pm0.010$ \\
		\bottomrule
	\end{tabular}
\end{table}

\subsection{Test for overfitting}\label{sec:overfitting}

To test for potential overfitting of our model, we tracked training and validation losses during training of our models. We did not find any stochastic significant difference between the two losses, indicating no sever memorization. Additionally, in \cref{tab:compiler_comparison_test_vs_train}, we compare the synthesis performance on unitaries of the train and test set, confirming good generalization.

\begin{table*}[!ht]
	\caption{\textbf{Test set vs. train set performance.} Reported values are averages over random 480 unitaries (from 2 to 16 gates), from the train or test set, and 128 circuits sampled for each unitary. Values in parenthesis are the corresponding standard deviation, showing the variability of the metrics.
    }
	\label{tab:compiler_comparison_test_vs_train}
	\centering
	\begin{tabular}{
			p{2.25cm}
			c
			>{\raggedleft\arraybackslash}p{2.85cm}
			>{\raggedleft\arraybackslash}p{1.7cm}
            >{\raggedleft\arraybackslash}p{2.2cm}
			>{\raggedleft\arraybackslash}p{1.9cm}
		}
		\toprule
        Method & \#~qubits & Avg. minimum infidelity & Avg. gate count & Avg. runtime per sample (seconds) & Avg. distinct circuits per unitary \\ 

        \midrule \midrule	
		genQC2 	    & 3 & $0.08\;(0.17)$ & $8\;(4)$ & $0.09\;(0.00)$ & $76\;(46)$ \\ 
		(train set) & 4 & $0.10\;(0.21)$ & $8\;(4)$ & $0.09\;(0.00)$ & $77\;(48)$ \\ 
		~ 			& 5 & $0.09\;(0.18)$ & $8\;(4)$ & $0.09\;(0.00)$ & $75\;(50)$ \\ 
        	
        \midrule	
		genQC2 	   & 3 & $0.09\;(0.20)$ & $8\;(4)$ & $0.09\;(0.00)$ & $78\;(45)$ \\ 
		(test set) & 4 & $0.10\;(0.20)$ & $8\;(4)$ & $0.09\;(0.00)$ & $74\;(49)$ \\ 
		~ 		   & 5 & $0.09\;(0.17)$ & $8\;(4)$ & $0.09\;(0.00)$ & $71\;(50)$ \\ 

		\bottomrule
	\end{tabular}
\end{table*}

\section{Dataset details}\label{sec:app_dataset}

In this section, we present the details of the training dataset used to train the model in \cref{sec:experiments}.

\paragraph{Dataset generation}
We create the dataset by generating random circuits consisting of the gates $\brek{\mathrm{h},\mathrm{cx},\mathrm{ccx},\mathrm{swap},\mathrm{rx},\mathrm{ry},\mathrm{rz},\mathrm{cp}}$ with CUDA-Q~\cite{cudaq} (see details on the latter below). For this, we first sample a subset of the gates from the previous set, and then append these gates  uniformly to an empty circuit. We do this randomly from 4 to 32 gates and for 3 to 5 qubits. Once the gate type is fixed, we sample the continuous parameters, of the parameterized gates, following a uniform distribution. After a circuit is sampled, we evaluate the implemented unitary and store the circuit-unitary pair. Additionally, we parse the sampled gate subset into a text prompt and store it alongside the circuits.

We sample random circuits until our dataset consists of 11 million (M) unique circuit ansätze. Sequentially, we expand the dataset by resampling the continuous parameters of circuits consisting of at least one parameterized gate six times. Note that this step means we have multiple copies of the same circuits in the dataset, but each with different parameters. With this method we increased the dataset size of 11M unique circuits to 63M unitaries, which we then use to train the model presented in \cref{sec:experiments}. Our final dataset has a memory size of $\sim$290 GB in total. Notably, to reduce training time and storage space, we store the unitaries in half precision (float16), which introduces a mean absolute error of $\mathrm{MAE}(UU^\dagger, I)\approx10^{-5}$ into the unitary property $UU^\dagger=I$. We assume that this is the potential reason of why the infidelities presented in \cref{fig:compile_rnd}b are mostly distributed above $10^{-5}$. Future works can explore training with higher precision unitaries.

\paragraph{Circuit simulation}

The dataset creation and verification was done via CUDA-Q~\cite{cudaq}, an open-source QPU-agnostic platform designed for accelerated quantum supercomputing. By offering a unified programming model in Python and C++ for co-located GPUs, QPUs, and CPUs, CUDA-Q enables the integration of classical and quantum resources within a single application, ensuring optimal performance and efficiency. The platform includes the NVQ++ compiler, which supports split compilation by lowering quantum kernels into multi-level intermediate representation (MLIR) and quantum intermediate representation (QIR). This approach ensures tight coupling between classical and quantum operations, facilitating accelerated execution of large-scale quantum workloads.

CUDA-Q's circuit simulation engine leverages NVIDIA's cuQuantum library, which supports state vector, density matrix, and tensor network simulations, enabling scaling to supercomputing scales. Users can with it switch the execution of their code from simulation to QPU hardware consisting of a rich variety of ionic, superconducting, photonic, neutral atoms and others as their hardware roadmap mature. Moreover, all executables have parallelization built into their functionality, hence execution of quantum kernels can be parallelized across multi-GPU architectures today, and multi-QPU architectures in the future. Recent additions to the platform have included specialized libraries for quantum error correction and quantum algorithm solvers, interoperability with the broader CUDA ecosystem and AI software, and cloud-based hardware access via services like Amazon Braket. 

\section{Training details}
\label{sec:app_training_details}

We train the Circuit-Diffusion-Transformer (CirDiT) (see \cref{sec:app_model_arch}) from \cref{sec:experiments} with the dataset detailed in \cref{sec:app_dataset}. For this, we optimize the loss defined in \cref{eq:loss} using the \textit{Adam} optimizer ~\cite{kingma2014adam}, with a learning rate of $10^{-4}$, $\beta_1=0.9$ and $\beta_2=0.999$, together with a one-cycle learning rate strategy~\cite{smith2019super}. Training is performed on 16 NVIDIA A100 GPUs for $\sim$800k update steps, with an effective batch size of 2048. The training time is roughly 700 single GPU hours. Additionally, as mentioned in \cref{sec:multidiff}, we drop the condition $\vb{c}$ with a probability of $10\%$, following classifier-free guidance (CFG)~\cite{ho2022classifier}.

For the multimodal diffusion process, we use for both modes the same number of times steps $T_h=T_w=T=1000$ (see details in \cref{sec:app_multimode_details}). We use for the discrete mode a learned discrete noise schedule for the linear target (see \cref{fig:DM_schedule}ac and \cref{sec:app_learned_schedule}) and the cosine squared schedule for the continuous mode (see \cref{fig:DM_schedule}bc and \cref{sec:app_cont_emb}). 

During training, to reduce the variance of the ELBO estimation, we sample the timesteps $t$ and $\tilde{t}$ of \cref{eq:loss} using a low-discrepancy sampling method, as done in Refs.~\cite{sahoo2024simple, kingma2021variational}. Considering a batch size of $m$, we sample both diffusion times for each batch sample $i$ by
\begin{equation}
	t_i, \tilde{t}_i \sim \mathcal{U}\bre{\frac{i-1}{m},\frac{i}{m}}\quad\text{for}\quad i\in\bre{1, \dots, m}.
\end{equation}
As the loss in \cref{eq:loss} requires two independent times, i.e. $t$ and $\tilde{t}$, we shuffle $\tilde{t}$ across the batch samples after sampling them. Furthermore, to increase the training time spend on similar times $t\approx\tilde{t}$, we do not shuffle the time step bins of $\tilde{t}$ with a probability of $5\%$, leaving $t$ and $\tilde{t}$ close together as they are sampled from the same bins, which means they differ by a maximal value of $1/m$.

\section{Inference details}

We sample the results from \cref{sec:experiments} using the joint sampling model, as explained in \cref{sec:app_sample_modes}, for 40 time steps. As diffusion sampler, we use the CFG++~\cite{chung2024cfgPP} variant of DPM++2M~\cite{lu2022dpm}. 

The benchmark runtime values, reported in \cref{tab:compiler_comparison}, are computed on a workstation with the \textit{AMD Ryzen Threadripper PRO 7955WX} CPU and a \textit{NVIDIA RTX PRO 6000 Blackwell Max-Q} GPU.

\subsection{Balanced testset} \label{sec:app_testset}

In \cref{sec:benchmark}, we showed the compilation of random unitaries from a test set. This test set is split from the training dataset (detailed in \cref{sec:app_dataset}), ensuring that the test unitaries are indeed compilable with the available gate set. Importantly, we do this before starting training and before resampling the parameters of the parametrized gates (see \cref{sec:app_dataset}), guaranteeing a benchmark set of unitaries resulting from unique circuits the model has never seen during training. Additionally, we balance the test set such that we have an equal amount of circuits per gate count.

\subsection{Multimodal CFG}

As explained in \cref{sec:multidiff}, the reverse transitions for each of the modes (see \cref{eq:split_gen_model}) have two conditions: one related to the opposite mode, and another one accounting for the external condition $\vb{c}$. For sampling using CFG~\cite{ho2022classifier}, we extend the guidance to two conditions.

Considering the guidance scales $\gamma_h, \gamma_w$ and $\lambda_h, \lambda_w$, we can write the guided velocity prediction (see \cref{sec:app_v_weight}) for the discrete mode as
\begin{align}
	\tilde{\vb{v}}_\theta^h(\vb{h}_{t}, \vb{w}_{\tilde{t}}, t, \tilde{t}, \vb{c})
	&= 
	\vb{v}_\theta^h(\vb{h}_{t}, \vb*{\epsilon}_w, t, 1, \phi) \nonumber\\&\quad+
	\gamma_h \bre{\vb{v}_\theta^h(\vb{h}_{t}, \vb{w}_{\tilde{t}}, t, \tilde{t}, \phi) - \vb{v}_\theta^h(\vb{h}_{t}, \vb*{\epsilon}_w, t, 1, \phi)} \nonumber\\&\quad+
	\lambda_h \bre{\vb{v}_\theta^h(\vb{h}_{t}, \vb{w}_{\tilde{t}}, t, \tilde{t}, \vb{c}) - \vb{v}_\theta^h(\vb{h}_{t}, \vb{w}_{\tilde{t}}, t, \tilde{t}, \phi)},
\end{align}
Analogously, for the continuous mode we write
\begin{align}
	\tilde{\vb{v}}_\theta^w(\vb{h}_{t}, \vb{w}_{\tilde{t}}, t, \tilde{t}, \vb{c})
	&= 
	\vb{v}_\theta^w(\vb*{\epsilon}_h, \vb{w}_{t}, 1, \tilde{t}, \phi) \nonumber\\&\quad+
	\gamma_w \bre{\vb{v}_\theta^w(\vb{h}_{t}, \vb{w}_{\tilde{t}}, t, \tilde{t}, \phi) - \vb{v}_\theta^w(\vb*{\epsilon}_h, \vb{w}_{t}, 1, \tilde{t}, \phi)} \nonumber\\&\quad+
	\lambda_w \bre{\vb{v}_\theta^w(\vb{h}_{t}, \vb{w}_{\tilde{t}}, t, \tilde{t}, \vb{c}) - \vb{v}_\theta^w(\vb{h}_{t}, \vb{w}_{\tilde{t}}, t, \tilde{t}, \phi)},
\end{align}
where $\vb*{\epsilon}_h, \vb*{\epsilon}_w \sim \mathcal{N}(\vb{0}, \vb{I})$ and $\phi$ represents an empty condition. In practice, we implement $\phi$ as a trainable latent vector. Empirically, we found the values $\gamma_h = 0.3$, $\gamma_w = 0.1$, $\lambda_h = 1.0$ and $\lambda_w = 0.35$ work well for circuit generation, which we then use for the results presented in \cref{sec:experiments}.

\section{Model architecture}\label{sec:app_model_arch}

\begin{figure}
    \centering
    \includegraphics[width=0.95\columnwidth]{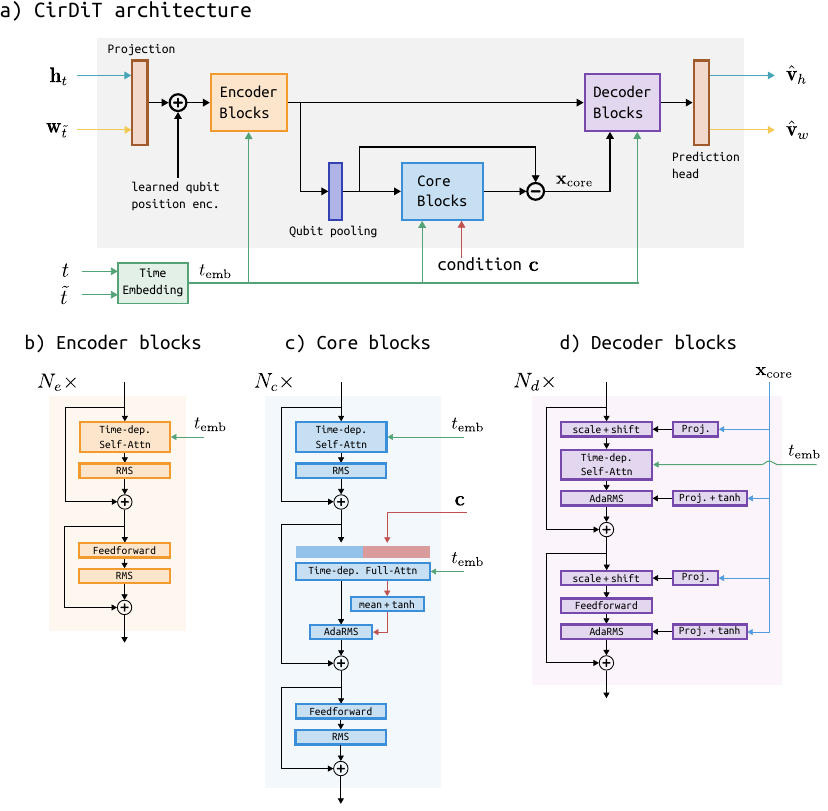}
    \caption{\textbf{Circuit Diffusion Transformer (CirDiT) architecture.}
    	\textbf{a)} Overview of the encoder-core-decoder structure. After passing the inputs through the encoder, the latent state is averaged across the qubit dimension, yielding a sequence of latent vectors. This sequence is then passed into the core transformer blocks, which inject the condition $\vb{c}$. Finally, the core output is passed alongside a skip connection of the encoder to the decoder.
    	\textbf{b)} Design of the encoder blocks.
    	\textbf{c)} Design of the core blocks. The condition $\vb{c}$ is passed through a multimodal full attention layer, where we concatenate the output of the previous layer with $\vb{c}$. Then, the output is split and the $\vb{c}$ portion is averaged and used as gating for the following AdaRMS layer.
    	\textbf{d)} Design of the decoder blocks. The core output is utilized by scaling and shifting operations on the main branch.
    	}\label{fig:app_cirdit}
\end{figure}

In \cref{fig:app_cirdit} we present the model architecture considered in this work, named Circuit-Diffusion-Transformer (CirDiT). Our model choice is based on the diffusion transformer (DiT) architecture~\cite{peebles2023scalable}, and contains 151 million trainable parameters. For the encoder and decoder blocks, we use a channel size of 256, a number of blocks $N_e=N_d=6$ and a number of 8 heads. The core transformer has 768 channels, $N_c=12$ blocks and 16 heads.

We inject the diffusion times $t$ and $\tilde{t}$ using Time-dependent Self-Attention (TMSA)~\cite{hatamizadeh2024diffit}, using a $t_{\text{emb}}$ size of 512 channels. The time dimension of the circuit embeddings (see \cref{sec:app_emb}) is encoded using the rotational position encoding $p$-RoPE of Ref.~\cite{barbero2025round}, setting the parameter of the latter to $p=0.9$. In addition, for the qubit dimension, we add a learned position encoding before the encoder blocks. As visualized in \cref{fig:app_cirdit}, we replace the Layer-Norm layers with RMS-Norms~\cite{zhang2019root} and push them to the end of the blocks, mitigating the divergence problem of large scale DiTs, as discussed in ~\cite{pmlr-v119-xiong20b, zhuo2024lumina}. We use in some layers the adaptive version AdaRMS-Norm, where we scale the normalized output by an external signal which is constrained by a $\mathrm{tanh}$ activation~\cite{zhuo2024lumina}.

\section{Unitary encoder pre-training}\label{sec:app_unitary_clip}

As explained in \cref{sec:multidiff}, in order to create an appropriate unitary conditioning, we pre-train a unitary encoder which encodes a given unitary $U$ together with a prompt into a condition $\vb{c}$. The main goal of this encoding is to align the representation of a given unitary to its circuit representation, ensuring that the information passed to the DM is as expressive as possible given the generation task.  

\begin{figure}
	\centering
	\includegraphics[width=0.95\columnwidth]{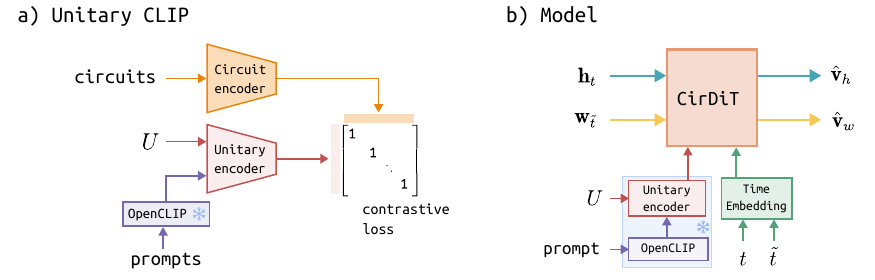}
	\caption{\textbf{UnitaryCLIP.}
		\textbf{a)} Overview of the contrastive unitary encoder pre-training.
		\textbf{b)} Conditioning of the diffusion model using the output of the unitary-prompt encoder as condition $\vb{c}$.
	}\label{fig:app_Uclip}
\end{figure}

Inspired by the contrastive loss of the CLIP framework~\cite{clip}, we match the latent encodings of a circuit encoder and a unitary-prompt encoder, as presented in \cref{fig:app_Uclip}a. Additionally, we first encode the text prompts using a frozen pre-trained \textit{OpenCLIP}~\cite{ilharcoOpenCLIP} model, specifically the architecture \textit{ViT-B-32} trained on the dataset \textit{datacomp\_xl\_s13b\_b90k}. The circuit and unitary encoder are based on the diffusion transformer architecture~\cite{peebles2023scalable}. The whole UnitaryCLIP (both encoders) we use in \cref{sec:experiments} contains 38 million trainable parameters. We use an additive absolute position encoding~\cite{Attention} for the unitary matrix elements.

We optimize the UnitaryCLIP using the \textit{Adam} optimizer ~\cite{kingma2014adam}, with a learning rate of $3.2\cdot10^{-4}$, $\beta_1=0.9$ and $\beta_2=0.999$, together with a one-cycle learning rate strategy~\cite{smith2019super}. We train for $\sim 550$k updates steps, with an effective batch size of 4096, on the dataset of \cref{sec:app_dataset}. We find that the trained UnitaryCLIP is able to achieve a 99\% correct matching of unseen unitary-circuit pairs.

\section{Additional figure parameters}

We list in \cref{tab:figure_parameters} additional sample parameters of the figures presented throughout this paper.

\begin{table}
    \centering
	\caption{Additional figure and table sampling parameters.}
	\label{tab:figure_parameters}
    \begin{tabular}{
    l
    >{\raggedleft\arraybackslash}p{1.5cm}
    >{\raggedleft\arraybackslash}p{2.3cm}
    >{\raggedleft\arraybackslash}p{6.0cm}
    }
    	\toprule
        ~  & Number of unitaries & Circuit samples per unitary &  Notes \\ 
        \midrule
        \cref{fig:compile_rnd}ab        & 1024	& 128	& \# unitaries per qubit count\\ 
        \cref{fig:compile_rnd}c         & 2048	& 128	& \# unitaries per qubit count\\ 
        \cref{fig:ansatz_fixing}        & 480	& 1	& \# unitaries per qubit count\\ 
        \cref{fig:model_ablation_scale} & 480	& 64 & \# unitaries per qubit count\\ 
        \cref{fig:hamiltonians}ab       & - & 128	& A unitary per grid point \\
        \cref{fig:hamiltonians}cd       & - & 128	& A unitary per time step \\
        \cref{fig:discovery}abc         & 1 & 2048	& QFT for 4 qubits  \\
        \cref{fig:discovery}defghi      & - & -	& Same circuits as in \cref{fig:hamiltonians}ab for 4 qubits \\
        \cref{fig:tree_search_paths}    & - & - & Same data as \cref{fig:ansatz_fixing} \\
        \cref{fig:app_qc_corruption}    & 16384 & 1	& Here \# unitaries is \# ciruits  \\
        \midrule
        \cref{tab:compiler_comparison}          & 480 & 128	& \# unitaries per qubit count\\ 
        \cref{tab:ablation_discrete_schedule}   & 480 & 64	& \# unitaries per qubit count \\ 
        \cref{tab:ablation_continuous_schedule} & 480 & 64	& \# unitaries per qubit count \\ 
        \bottomrule              
    \end{tabular}   
\end{table}

\section{Circuit corruption test}\label{sec:app_circuit_corruption}

\begin{figure}
	\centering
	\includegraphics[width=0.9\columnwidth]{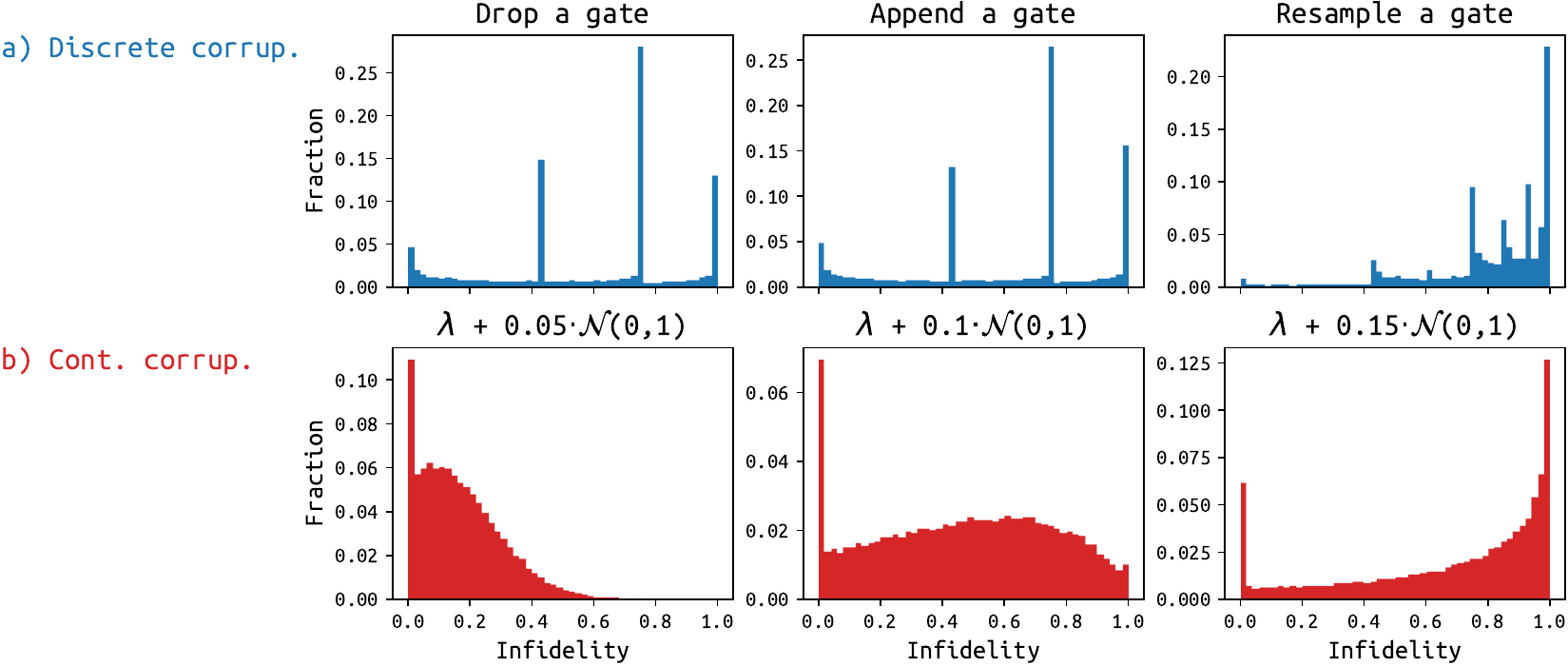}
	\caption{\textbf{Circuit corruptions.}
		\textbf{a)} Infidelity distribution for circuits with discrete corruptions (see \cref{sec:app_circuit_corruption}).
		\textbf{b)} Infidelity distribution for circuits with continuous corruption, as defined in \cref{eq:app_cont_corrup}.
		\label{fig:app_qc_corruption}
	}
\end{figure}

In \cref{fig:compile_rnd}a we observed the appearance of some characteristic peaks in the infidelity distribution. In order to further investigate the origin of such peaks, we perform different corruptions on test set circuits and record the resulting infidelities between corrupted and original circuits. We show in \cref{fig:app_qc_corruption} the infidelities for the following discrete and continuous corruptions:
\begin{enumerate}
	\item Drop a single random gate from the circuit.
	\item Append a single random gate from the gate set, with random connections.
	\item Replace a single random gate with a random one from the gate set, with random connections.
	\item Add Gaussian noise to the normalized parameters $\lambda\in[-1,1]$ of all continuous gates in a circuit, i.e.
	\begin{equation}\label{eq:app_cont_corrup}
		\tilde{\lambda}=\lambda + A \cdot \mathcal{N}(0,1),
	\end{equation}
	for $A\in\brek{0.05, 0.1, 0.15}$ (see \cref{fig:app_qc_corruption}b).
\end{enumerate}
Note, we always take only a single corruption per circuit from the list above.

Comparing the peaks in the discrete corruptions of \cref{fig:app_qc_corruption}a to the ones observed in the random unitary compilation (see \cref{sec:benchmark}), we see the same peaks appearing at $0.4$ and $0.8$. Hence, we attribute the peaks arising in \cref{fig:compile_rnd}a to cases in which the model incorrectly places gates in a similar way as our corruption tests, this means, by misplacing a single gate. On the other hand, corrupting the continuous values of the parameterized gates causes a wide continuous distribution around zero (\cref{fig:app_qc_corruption}b), where the peak at zero infidelity corresponds to circuits which have no parameterized gates. This highlights an important result: comparing this distribution to that in \cref{fig:compile_rnd}a showcases the fact that the model predicts the continouous parameters rather accurately, as no broad distribution arises, and the error arise mainly from misplacements of a single gate.

\section{Noisy simulation}\label{sec:app_noisy_sim}

To test the noise robustness of the synthesized circuits, we simulate them under the influence of gate errors. We then calculate the process infidelity as a measure to estimate how well the circuits could run on actual NISQ hardware, as we show in \cref{fig:noisy_simulation}.

In the noiseless simulation, we can write the channel of a target unitary $U$ as
\begin{equation}
    \mathcal{E}_{U}(\rho) = U\rho U^\dagger,
\end{equation}
where $\rho$ is the density matrix. Next, we define the process infidelity--a distance between the target channel $\mathcal{E}_{U}$ and the noisy channel $\mathcal{E}_{\text{noisy}}$, which is given by a synthesized circuit in the presence of gate errors--as
\begin{equation}
    \mathcal{I}(\mathcal{E}_{\text{noisy}}, \mathcal{E}_{U}) = 1 - \frac{1}{d^2} \mathrm{Tr} \left[  S_{\mathcal{E}_{U}}^\dagger \mathcal{S}_{\mathcal{E}_{\text{noisy}}} \right],
\end{equation}
where $d$ is the dimension of the channels. Here, $S_{\mathcal{E}_{U}}$ and $\mathcal{S}_{\mathcal{E}_{\text{noisy}}}$ are the super-operator representations of the channels.

As a simplified model, we describe the noisy circuits with depolarizing errors applied to all gates. Specifically, we assign each $k$-qubit gate, with unitary $U_{\text{gate}}=U_k \otimes \vb{I}$ and connections on the qubits $\{ k_i\}$, placed in a $n$-qubit circuit, a depolarizing error given by
\begin{equation}
    \mathcal{E}_k(\rho) = (1-p_k)\;U_{\text{gate}} \rho U_{\text{gate}}^\dagger + p_k\; \frac{1}{2^k} \vb{I}_{\{ k_i\}} \otimes \mathrm{Tr}_{\{k_i\}}[\rho] ,
\end{equation}
where the error acts simultaneously on the $k$ qubits. Here, $\mathrm{Tr}_{\{k_i\}}$ is the trace over all $k$ qubits the gate acts on, $\vb{I}_{\{ k_i\}}$ is the corresponding identity matrix, and $p_k$ is the probability that the depolarizing error occurs.

To parametrize the error probabilities $p_k$ with a single control parameter $p$, we model the probability that a $k$-qubit gate acts without error as the probability that $k$ independent single qubit gates act without an error, i.e. we set
\begin{equation}
    1-p_k \overset{!}{=} (1-p_1)^k.
\end{equation}
This relation then results in the equations:
\begin{equation}
    p_1 = p, \quad p_2 = 1-(1-p)^2 \text{ and}\quad p_3 = 1-(1-p)^3,
\end{equation}
which we use for the results shown in \cref{fig:noisy_simulation}.

\section{Hamiltonians}\label{sec:app_ham}

In \cref{sec:hamiltonian} we showed the compilation of the evolution operators of the Ising and XXZ Hamiltonians. Here we define these Hamiltonians in terms of the Pauli operators, which are defined as
\begin{equation}    
X = \begin{pmatrix}0 & 1 \\ 1 & 0\end{pmatrix},\;
Y = \begin{pmatrix}0 & -i \\ i & 0\end{pmatrix}\;\text{and}\;
Z = \begin{pmatrix}1 & 0 \\ 0 & -1\end{pmatrix}.
\end{equation}
For both Hamiltonians, we consider here the case in which the $n$ qubits of the quantum circuit represent the spins of a non-periodic one-dimensional chain, where we write neighboring spins $i$ and $j$ as ${\langle i, j \rangle}$.

\paragraph{Ising Hamiltonian:}
Defined as
\begin{equation}\label{eq:app_ising_h}
	H_\text{ising} = -J \sum_{\langle i, j \rangle} Z_i Z_j - h \sum_{i=0}^{n-1} X_i,
\end{equation}
where $J\in\reals$ is the coupling constant and $h\in\reals$ a magnetic field.

\paragraph{XXZ Hamiltonian:}
Defined as
\begin{equation}\label{eq:app_xxz_h}
	H_\text{xxz} = -J \sum_{\langle i, j \rangle} ( X_i X_j + Y_i Y_j + \Delta Z_i Z_j ) - h \sum_{i=0}^{n-1} X_i,
\end{equation}
where $J\in\reals$ is the coupling constant, $\Delta\in\reals$ a perturbation and $h\in\reals$ a magnetic field. In \cref{sec:experiments}, we fix $h=0.2$ for the XXZ Hamitonian.

\section{Additional GPE structures} \label{sec:app_add_gadgets}

In this section, we present additional circuit structures extracted using our proposed Gate-Pair Encoding (GPE) scheme (see \cref{sec:gpe}), extending the results presented in \cref{sec:discovery}. We follow the exact same recipe as in the main text, and take all circuits generated by the DM, without any selection or filtering for wrong circuits. Notably, this means all structures are extracted without \emph{any} circuit evaluation, keeping all computations of the process purely classic.

\begin{figure}
	\centering
	\includegraphics[width=0.9\columnwidth]{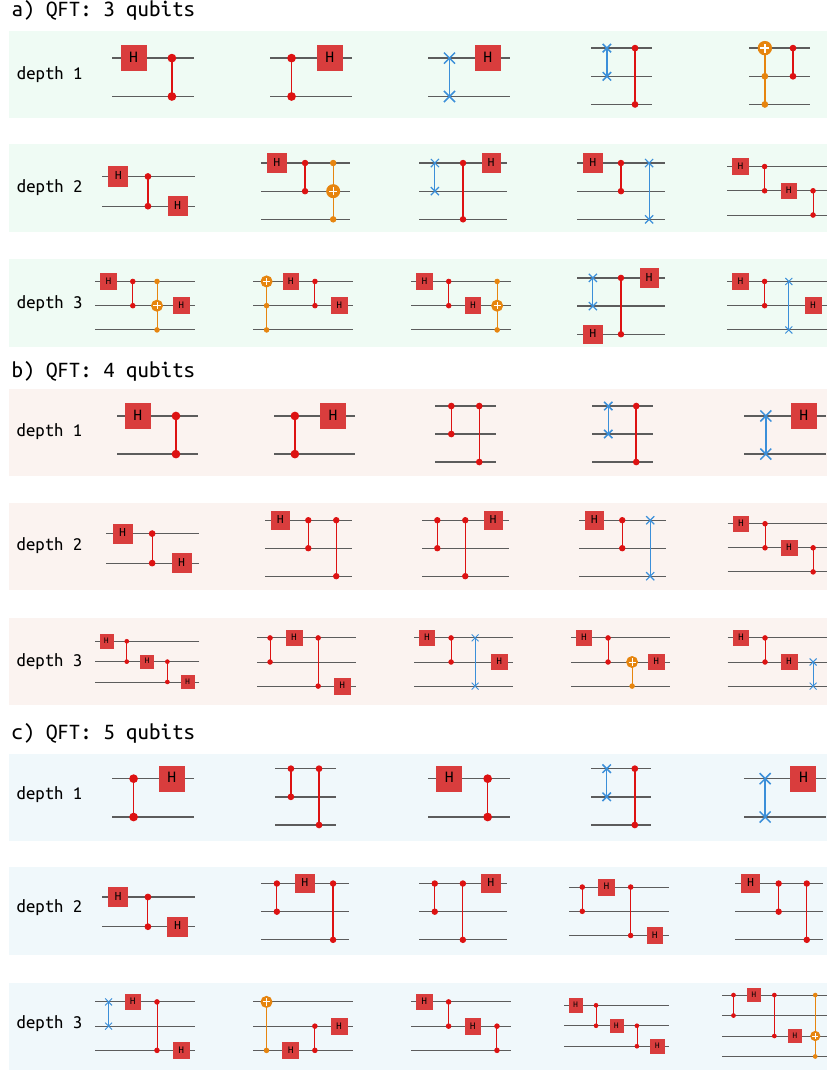}
	\caption{\textbf{QFT Gate-Pair Encodings of genqc2.}
		 GPE (see \cref{sec:gpe}) on generated circuits for the QFT unitary for \textbf{a)} 3 qubits, \textbf{b)} 4 qubits and \textbf{c)} 5 qubits.
		The rows correspond to the token depth (see \cref{sec:app_add_gadgets}). The circuits shown represent the top-5 most occurring structures, from most occurring (left) to less often occurring (right).
	}
	\label{fig:app_additional_gadgets_qft}
\end{figure}

\begin{figure}
	\centering
	\includegraphics[width=0.9\columnwidth]{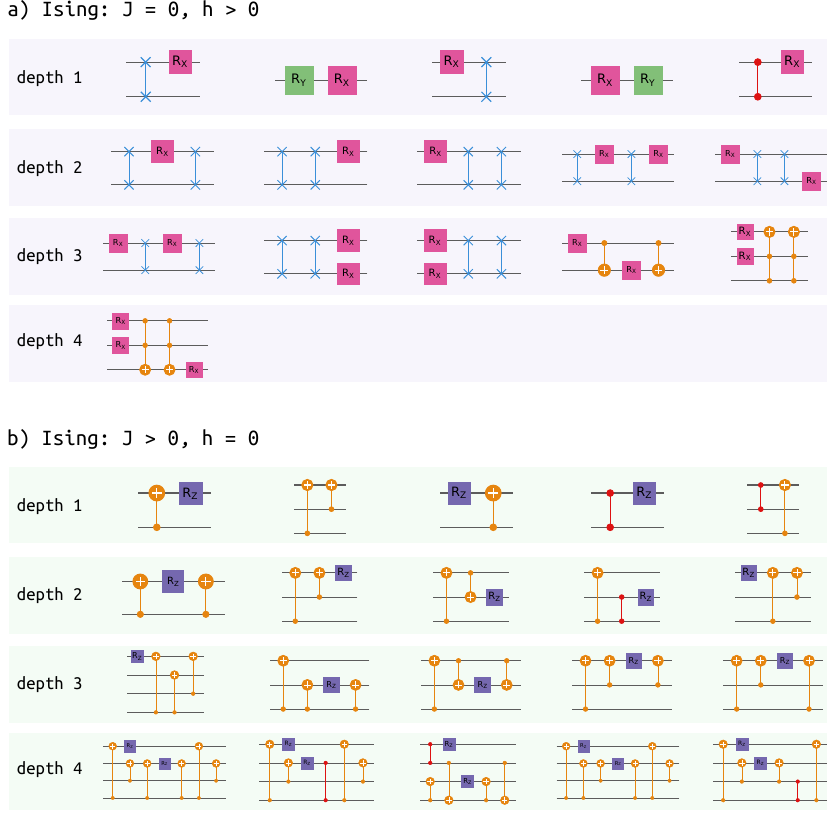}
	\caption{\textbf{Ising Hamiltonian Gate-Pair Encodings of genqc2.} 
		 GPE (see \cref{sec:gpe}) on generated circuits for the 4-qubit Ising Hamiltonian (defined in \cref{eq:app_ising_h}) evolution unitary for $\tau=0.25$.
		We use in \textbf{a)} circuits for the parameters $h\in\bre{0.5, 0.9}$ and $J=0$, and in \textbf{b)} $J\in\bre{0.5, 0.9}$ and $h=0$.	
		The rows correspond to the token depth (see \cref{sec:app_add_gadgets}). The circuits shown represent the top-5 most occurring structures, from most occurring (left) to less often occurring (right).	
		}
	\label{fig:app_additional_gadgets_ising}
\end{figure}

\begin{figure}
	\centering
	\includegraphics[width=0.9\columnwidth]{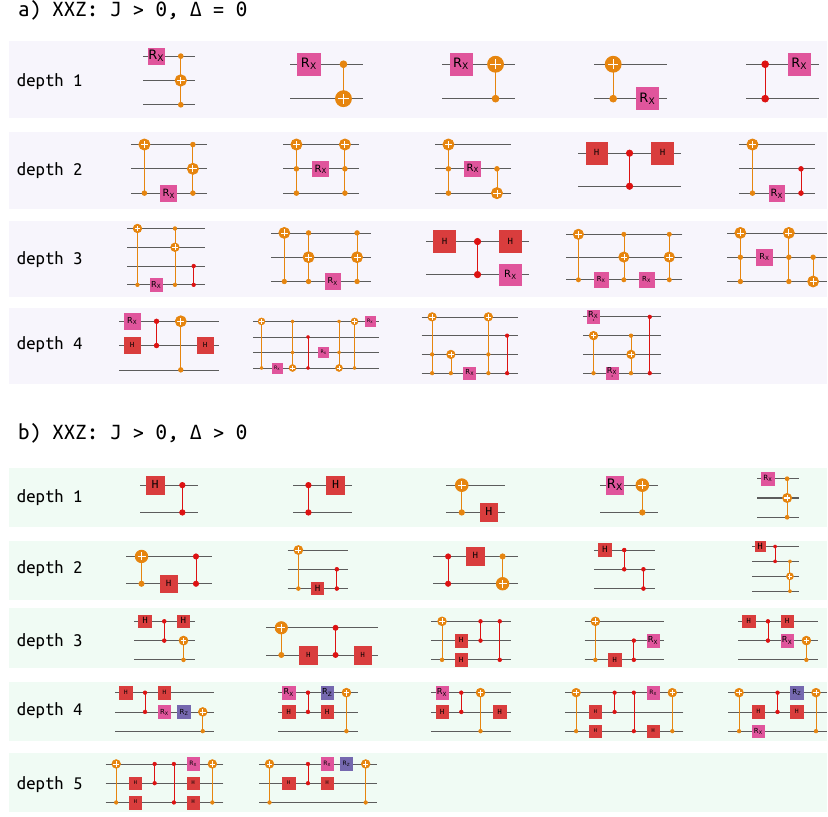}
	\caption{\textbf{XXZ Hamiltonian Gate-Pair Encodings of genqc2.} 
		 GPE (see \cref{sec:gpe}) on generated circuits for the 4-qubit XXZ Hamiltonian (defined in \cref{eq:app_xxz_h}) evolution unitary for $\tau=0.25$.
		We use in \textbf{a)} circuits for the parameters $J\in\bre{0.5, 0.9}$ and $\Delta = 0$, and in \textbf{b)} $J\in\bre{0.5, 0.9}$ and $\Delta\in\bre{0.5, 0.9}$.	
		The rows correspond to the token depth (see \cref{sec:app_add_gadgets}). The circuits shown represent the top-5 most occurring structures, from most occurring (left) to less often occurring (right).
	}
	\label{fig:app_additional_gadgets_xxz}
\end{figure}

\begin{figure}
	\centering
	\includegraphics[width=0.9\columnwidth]{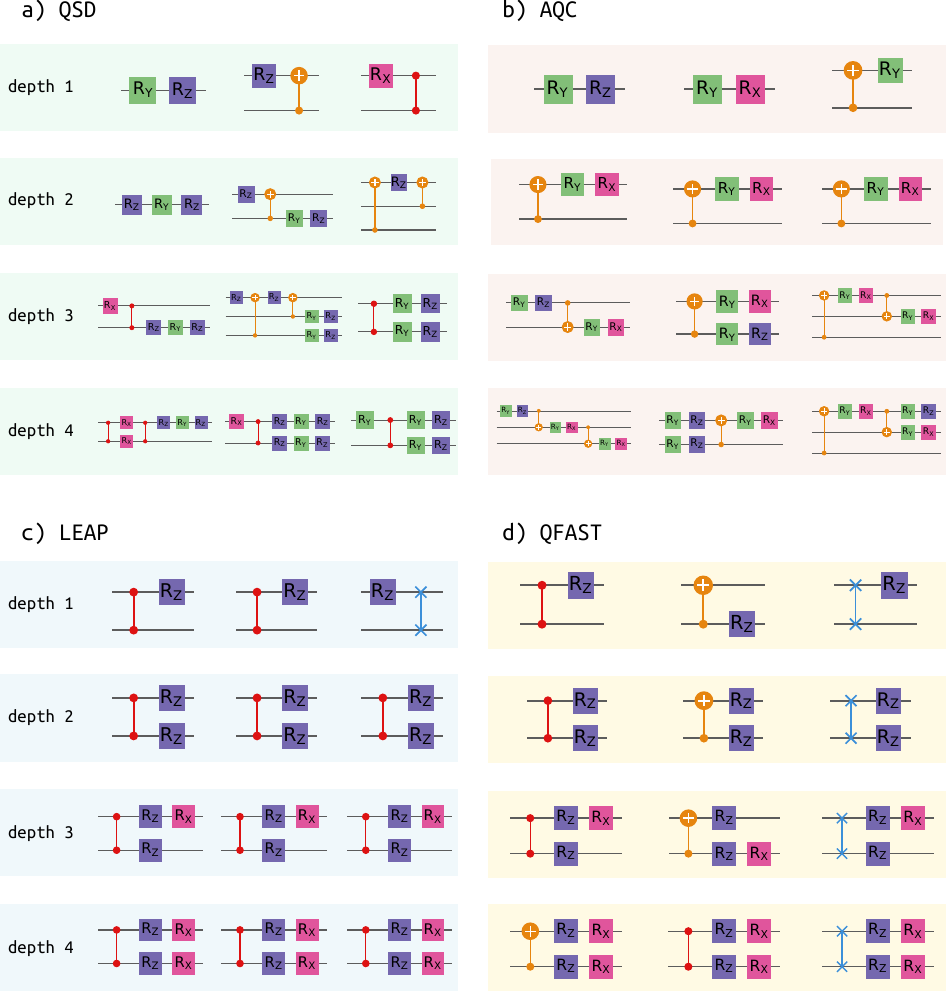}
    \caption{\textbf{QFT Gate-Pair Encodings.}
		 GPE (see \cref{sec:gpe}) on generated circuits for the 4-qubit QFT unitary for \textbf{a)} QSD, \textbf{b)} AQC, \textbf{c)} LEAP and \textbf{d)} QFAST.
		The rows correspond to the token depth (see \cref{sec:app_add_gadgets}). The circuits shown represent the top-3 most occurring structures, from most occurring (left) to less often occurring (right).
	}
	\label{fig:app_additional_gadgets_qft_other_methods}    
\end{figure}

For the structures shown in \cref{fig:app_additional_gadgets_qft}, \cref{fig:app_additional_gadgets_ising}, \cref{fig:app_additional_gadgets_xxz} and \cref{fig:app_additional_gadgets_qft_other_methods},
we run GPE for a maximum of 250 iterations, or until there are no pairs left to be matched into a new token. We present structures for different token depths, where depth 0 is defined as the elementary tokens, being here the original gate set (e.g. $\mathrm{h}$ or $\mathrm{cx}$). Then, depth 1 structures are gate-pairs (i.e. pairs of depth 0 tokens). Further, depth 2 tokens are constructed from at least one depth 1 token together with either one depth 1 or 0 token. More generally, a depth $m$ token always consists of one depth $m-1$ token and another one with depth $\leq m-1$.

We present in \cref{fig:app_additional_gadgets_qft}, \cref{fig:app_additional_gadgets_ising} and \cref{fig:app_additional_gadgets_xxz} structures generated by our method (genqc2), compared to structures from other methods in \cref{fig:app_additional_gadgets_qft_other_methods}.


